\documentclass[lettersize,journal]{IEEEtran}
\usepackage{amsmath,amsfonts}
\usepackage{algorithmic}
\usepackage{algorithm}
\usepackage{array}
\usepackage[caption=false,font=normalsize,labelfont=sf,textfont=sf]{subfig}
\usepackage{textcomp}
\usepackage{stfloats}
\usepackage{url}
\usepackage{verbatim}
\usepackage{graphicx}
\usepackage{cite}
\usepackage{booktabs}
\usepackage{multirow}
\usepackage{makecell}
\usepackage{tabularx}
\usepackage{float}
\usepackage{url}
\usepackage{hyperref}
\usepackage[numbers]{natbib}
\usepackage[utf8]{inputenc}

\usepackage{cleveref}  
\usepackage{marvosym}

\usepackage{siunitx}
\usepackage{graphicx}

\usepackage{amssymb}  
\begin{document}

\title{IP-Augmented Multi-Modal Malicious URL Detection Via
Token-Contrastive Representation Enhancement and Multi-Granularity Fusion}

\author{
Ye Tian, 
Yanqiu Yu, 
Liangliang Song,  
Zhiquan Liu, 
Yanbin Wang\textsuperscript{\Letter}, 
Jianguo Sun\textsuperscript{\Letter}

\thanks{

Yanbin Wang is with the Department of Engineering, Shenzhen MSU-BIT University, Shenzhen, China 518172.

Zhiquan Liu is with the School of Cyberspace Security, Jinan University, Guangzhou, China 510632.

Ye Tian, Yanqiu Yu, Liangliang Song, Jianguo Sun are with the Hangzhou Institute of Technology, Xidian University, Hangzhou, China 311200.
Ye Tian, Yanqiu Yu and Liangliang Song contributed equally to this work.

\textsuperscript{\Letter}Yanbin Wang and Jianguo Sun are the corresponding author.

}
}


\maketitle
\begin{abstract}
Malicious URL detection remains a critical cybersecurity challenge as adversaries increasingly employ sophisticated evasion techniques including obfuscation, character-level perturbations, and adversarial attacks. Although pre-trained language models (PLMs) like BERT have shown potential for URL analysis tasks, three limitations persist in current implementations: (1) inability to effectively model the non-natural hierarchical structure of URLs, (2) insufficient sensitivity to character-level obfuscation, and (3) lack of mechanisms to incorporate auxiliary network-level signals such as IP addresses — all essential for robust detection.

To address these challenges, we propose CURL-IP, an advanced multi-modal detection framework incorporating three key innovations: (1) Token-Contrastive Representation Enhancer, which enhances subword token representations through token-aware contrastive learning to produce more discriminative and isotropic embeddings; (2) Cross-Layer Multi-Scale Aggregator, employing hierarchical aggregation of Transformer outputs via convolutional operations and gated MLPs to capture both local and global semantic patterns across layers; and (3) Blockwise Multi-Modal Coupler that decomposes URL-IP features into localized block units and computes cross-modal attention weights at the block level, enabling fine-grained inter-modal interaction. This architecture enables simultaneous preservation of fine-grained lexical cues, contextual semantics, and integration of network-level signals. Our evaluation on large-scale real-world datasets shows the framework significantly outperforms state-of-the-art baselines across binary and multi-class classification tasks. The model also demonstrates particular strengths in: (1) robustness against compound adversarial attacks, (2) maintaining performance under class imbalance, and (3) practical effectiveness in real-world threat detection, validated through case studies. The code and datasets are publicly available at: \url{https://github.com/sevenolu7/MACFormer}.

\end{abstract}

\begin{IEEEkeywords}
Malicious URL Detection, Multi-modal Fusion, Character-aware Representation, Transformer
\end{IEEEkeywords}

\section{Introduction}
\IEEEPARstart Malicious URLs serve as digital traps, strategically crafted by attackers to exploit vulnerabilities in online behavior. These deceptive links often mimic legitimate websites, luring users into interactions that compromise their security. Beyond immediate financial losses, they erode trust in digital platforms and expose individuals to long-term privacy risks. The sophistication of these threats continues to evolve, employing advanced social engineering tactics to bypass traditional security measures, making them a persistent challenge in cybersecurity defense \cite{laszka2015optimal,bitaab2023beyond,guo2022safer}. In Q4 2024, APWG\cite{r3} recorded 989,123 phishing attacks, while the average wire transfer request in BEC scams surged to \$128,980—nearly double the previous quarter’s figure, underscoring the escalating financial impact of these threats.

\begin{figure}[!t]
\centering
\includegraphics[width=0.95\linewidth]{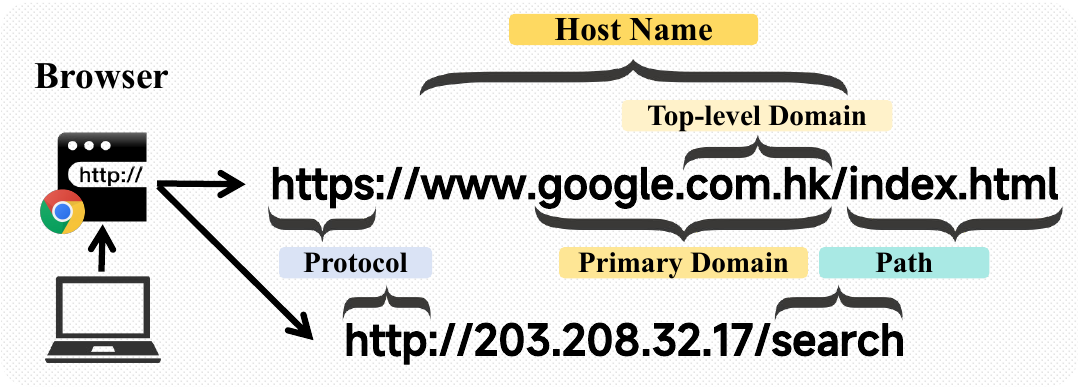}
\caption{Example of URL}
\label{fig_1}
\end{figure}

As cyber threats grow increasingly sophisticated, attackers frequently impersonate trusted brands like Microsoft and Google, exacerbating the challenges of malicious URL detection \cite{r4}. Traditional detection methods \cite{sahoo2017malicious}—including blacklists, heuristic analysis, and rule-based systems—remain constrained by their dependence on known URL structures and manual updates, leading to delayed and inconsistent responses to novel threats \cite{r6,r8,sabir2022reliability,lin2021phishpedia,zhang2011textual}. Early approaches that relied on manually extracted lexical or statistical features (e.g., URL length, dot frequency, or keywords) \cite{r9,r10} further demonstrated limited generalizability against dynamic attack patterns. These limitations highlight the imperative for advanced machine learning techniques that can autonomously exploit structural and lexical patterns (\Cref{fig_1}), enabling adaptive, precise, and real-time threat detection \cite{r8,r9,r10,liu2024less}.

Despite the notable success of Convolutional Neural Networks (CNNs) in malicious URL detection—demonstrated by state-of-the-art models such as URLNet\cite{r11}, TException\cite{r12}, and TextCNN\cite{r13}]—the inherent limitations of CNN-based architectures, including their restricted receptive fields and translation-invariant feature extraction mechanisms, are increasingly recognized as critical bottlenecks to further performance gains. Meanwhile, the rapid advancement of deep learning techniques has continuously driven progress in this field\cite{r1,r14,r15,r16,r17}, motivating exploration beyond traditional CNN frameworks\cite{tsai2024toward,sabir2022reliability,liang2021robust}.

While pre-trained language models (PLMs) such as BERT(Bidirectional Encoder Representations from Transformers)\cite{r18} have shown impressive success in natural language processing tasks\cite{r19,r20,r21}, their direct application to URL-based malicious detection remains limited. Existing PLMs often struggle with modeling the non-natural, hierarchical structure of URLs, lack sensitivity to character-level perturbations, and fail to effectively capture localized lexical patterns—factors that are crucial for identifying obfuscated or adversarial URLs\cite{cao2025phishagent,liu2022inferring}.

In addtional, traditional URL-only detection models depend entirely on lexical and syntactic patterns—features that adversaries can systematically evade through character-level perturbations \cite{r23}, homograph attacks \cite{r24}, or dynamic domain generation algorithms (DGAs)\cite{r22}. In contrast, IP address features provide complementary infrastructure-level signals with superior temporal stability, capturing three critical threat intelligence dimensions: (1) Reputation (malicious URLs frequently resolve to IPs with documented abuse histories), (2) Hosting Patterns (suspicious IPs often exhibit abnormal co-hosting behaviors, such as serving \textgreater 
100 phishing domains), and (3) Geospatial Anomalies (over 60\% of phishing infrastructure concentrates in specific ASNs/geolocations per APWG reports \cite{r3}). By fusing these orthogonal signals, our model effectively disambiguates semantically similar benign and malicious URLs—particularly in low-resource or ambiguous scenarios. Experimental results demonstrate that combining URL and IP modalities enhances detection robustness, especially under class imbalance and adversarial conditions.

Our proposed architecture introduces several key innovations to overcome these limitations, specifically designed to capture: (1) multi-level semantic representations, (2) character-aware signals, and (3) local-global interactions within URL sequences. Unlike conventional content-based URL analysis methods, we uniquely integrate IP address features as auxiliary contextual information. This cross-modal design enables simultaneous learning of network-level behavioral patterns and lexical semantics, providing complementary threat detection capabilities that are robust to obfuscation techniques.

\textbf{The main contributions are summarized as follows:}

\begin{itemize}
    \item We present CURL-IP, an IP-enhanced multimodal framework for malicious URL detection that addresses three fundamental limitations of existing PLM-based approaches: (1) insufficient lexical granularity, (2) inability to leverage hierarchical contextual cues, and (3) lack of mechanisms to incorporate auxiliary network signals like IP addresses.

    \item Our approach advances URL representation learning by combining Token-Aware Contrastive Learning with BERT (TACL-BERT) for discriminative, isotropic embeddings, complemented by knowledge distillation to preserve crucial semantic patterns across different URL structures.

    \item Our proposed Cross-Layer Multi-Scale Aggregator (CLMSA) module combines features from all hidden layers of the Transformer through a hierarchical fusion mechanism comprising cascaded convolutional blocks and gated multilayer perceptrons (gMLPs). The convolutional extracts local structural patterns characteristic of URL, while the gMLP preserve global contextual information. 

    \item Our Blockwise Multimodal Coupler (BMMC) module decomposes URL-IP embeddings into localized block units and computing block-level cross-modal attention weights, enabling fine-grained feature alignment between discrete URL tokens and continuous IP address features. 

    \item We extensively evaluate our method across diverse real-world threat scenarios—binary/multi-class classification, few-shot learning, adversarial attacks, and case studies—consistently outperforming strong baselines, achieving SOTA AUC and F1-scores in both standard and adversarial settings, proving its practical robustness.
\end{itemize}

\section{Related Work}
Malicious URL detection has been studied for a long time. We focus on recent advances most relevant to our work: (1) character-aware methods and (2) transformer-based approaches.

\subsection{Character-aware Methods}
Character-aware language models have demonstrated significant advantages in malicious URL detection tasks. By operating directly on raw character sequences instead of relying solely on fixed vocabularies, they are capable of capturing subtle lexical patterns such as special characters, obfuscation strategies, and misspellings. URLNet\cite{r11} pioneered a dual-channel CNN architecture to jointly learn character-level and word-level embeddings, which inspired many follow-up studies\cite{r28,r12,r29,r30} that adopted similar dual-input schemes. URLBERT\cite{r27} designed a Transformer-based encoder with a custom character tokenizer and employed contrastive learning to enhance structural understanding of URLs. Similarly, CGRU\cite{r25} combines character-level CNNs and GRU units to extract both local patterns and sequential dependencies, further aided by malicious keyword injection. Other methods such as the BiLSTM-based model\cite{r26} leverage bidirectional recurrence to capture long-range dependencies across character sequences, while fusing character and word-level information to better represent complex URL structures.To address the position sensitivity issue inherent in token-based models, Liu et al.\cite{r40} introduced a position-aware embedding mechanism that improves robustness to adversarial perturbations by encoding relative character positions.

Despite their progress, existing character-aware models often rely on dual-input architectures or shallow fusion strategies, which limit the integration of lexical and semantic signals. In contrast, we propose to adopt TACL-BERT to encode URL, which distills character-level morphological knowledge into subword token representations via a student–teacher training paradigm. This enables our model to preserve both character sensitivity and contextual expressiveness within a single Transformer backbone, avoiding additional tokenizer design or fusion overhead. Furthermore, our method stacks and fuses multi-layer representations hierarchically, capturing semantic patterns at different depths — an ability often missing in prior works. 

\subsection{Transformer-based Methods}
Transformer-based models have recently shown promising results in malicious URL detection due to their strong sequence modeling capabilities. Early works, such as Chang et al.\cite{r14} and URLTran\cite{r1}, fine-tuned general-purpose BERT models on URL datasets for phishing and malicious link detection. While these approaches demonstrated feasibility, they suffered from domain mismatch between the natural language pre-training corpus and the highly structured, non-linguistic format of URLs. To address this, Wang et al.\cite{r33} proposed training a BERT model from scratch on a large-scale URL dataset, thereby improving domain adaptation at the cost of high computational resources. Meanwhile, BERT-CNN\cite{r34} combined BERT embeddings with convolutional neural networks to capture both global context and local patterns, while CharBERT\cite{r35} introduced a dual-channel encoder to fuse subword and character-level features, along with a Noisy LM task to simulate adversarial perturbations.TransURL\cite{r38} built upon CharBERT by incorporating spatial pyramid attention, deep separable convolution, and multi-level encoding fusion to better capture local semantics.  Other domain-adaptive approaches such as DomURLs\_BERT\cite{r36} and PMANet\cite{r37} employed customized tokenizers, structure-aware inputs, and post-training strategies to enhance model robustness and structural alignment with URLs. URLBERT\cite{r27}, as the first PLM specifically designed for URL analysis, introduced contrastive learning, virtual adversarial training, and multi-task learning, achieving strong generalization across URL-related tasks.Recently, Zhang et al.\cite{r41} proposed using large language models (LLMs) to automatically generate high-quality semantic features, reducing the need for manual feature engineering as required by models like URLBERT.

Despite these advances, many of the above methods either rely on heavy architectural modifications, require extensive domain-specific pretraining, or employ separate modules for character-level and contextual learning. In contrast, our approach leverages a token-aware contrastive learning framework that injects character-level morphology into subword embeddings via a lightweight student–teacher distillation strategy. This avoids the need for dual-channel encoders or tokenizer customization while maintaining fine-grained lexical sensitivity. Furthermore, our model enhances feature expressiveness CLMSA module that aggregates multi-layer Transformer representations, and a BMMC module that integrates auxiliary IP address embeddings. 

\section{Dataset}
We construct three evaluation datasets by sampling from DeepURLBench, a public repository of labeled URLs with IP metadata. To ensure sample uniqueness, we employ non-overlapping sampling across datasets and analyze discriminative network features including top-level domain distributions, IP address class allocations (A-E), and top Autonomous System Numbers (ASNs). These features reveal significant infrastructural disparities between benign and malicious URLs. 

\textbf{URL-Binary (Binary Classification):}
The binary classification benchmark comprises 802,228 samples, with a near-balanced composition of 399,923 malicious and 402,305 benign URLs. As summarized in Table~\ref{tab:dataset_url_stats}, malicious and benign URLs differ substantially in infrastructural patterns, with malicious samples more frequently appearing in ccTLDs and less-regulated gTLDs. Furthermore, IP class distribution (Table~\ref{tab:ip_class}) shows malicious domains tend to cluster in Class A and C address ranges. ASN-level statistics (Table~\ref{tab:asn_dist}) reveal distinct hosting patterns between benign and malicious sources.

\textbf{URL-Adversarial (Adversarial Evaluation):}
This dataset evaluates model robustness through 160,000 samples (80,000 benign, 40,000 malicious, 40,000 adversarial), where adversarial samples are generated via evasive character injection to simulate obfuscation attacks. Compared to URL-Binary, these samples exhibit greater heterogeneity in TLD and IP class distributions, while \Cref{tab:asn_dist} demonstrates their less concentrated ASN patterns, significantly increasing detection complexity.

\textbf{URL-MultiClass (Multi-class Classification):} The dataset contains 671,957 URLs (520,631 benign, 30,182 malicious, 121,144 phishing) representing a realistic imbalanced threat distribution. While maintaining separate phishing and malicious classes for model evaluation, we consolidate them as "malicious" in Tables~\ref{tab:dataset_url_stats}--\ref{tab:asn_dist} for consistent comparison with binary classification results. This approach preserves the dataset's multi-class nature while enabling direct performance benchmarking across different threat detection scenarios.

\begin{table*}[htbp]
\centering
\caption{The Statistics of Our Dataset}
\label{tab:dataset_url_stats}
\resizebox{\textwidth}{!}{
\begin{tabular}{@{}l*{9}{l}@{}}
\toprule
\multirow{2}{*}{Dataset} & \multicolumn{3}{l}{Sample Sizes} & \multicolumn{3}{l}{Benign TLDs} & \multicolumn{3}{l}{Malicious TLDs} \\
\cmidrule(lr){2-4} \cmidrule(lr){5-7} \cmidrule(lr){8-10}
 & \multicolumn{1}{l}{malicious} & \multicolumn{1}{l}{benign} & \multicolumn{1}{l}{total} & \multicolumn{1}{l}{.com} & \multicolumn{1}{l}{ccTLDs} & \multicolumn{1}{l}{other gTLDs} & \multicolumn{1}{l}{.com} & \multicolumn{1}{l}{ccTLDs} & \multicolumn{1}{l}{other gTLDs} \\
\midrule
Dataset I & 399,923 & 402,305 & 802,228 & 47.79\% & 40.13\% & 12.08\% & 45.40\% & 30.71\% & 23.90\% \\
Dataset II & 80,000 & 80,000 & 160,000 & 52.39\% & 35.64\% & 11.96\% & 47.79\% & 20.57\% & 31.64\% \\
Dataset III & 151,326 & 520,631 & 671,957 & 53.05\% & 33.59\% & 13.37\% & 55.54\% & 24.58\% & 19.89\% \\
\bottomrule
\end{tabular}
}
\end{table*}

\begin{table*}[htbp]
\centering
\caption{IP Address Class Distribution}
\label{tab:ip_class}
\begin{tabular}{@{}l*{8}{l}@{}}
\toprule
\multirow{2}{*}{Dataset} & \multicolumn{4}{c}{Benign IP Class Distribution} & \multicolumn{4}{c}{Malicious IP Class Distribution} \\
\cmidrule(lr){2-5} \cmidrule(lr){6-9}
 & \multicolumn{1}{l}{A} & \multicolumn{1}{l}{B} & \multicolumn{1}{l}{C} & \multicolumn{1}{l}{D/E} & \multicolumn{1}{l}{A} & \multicolumn{1}{l}{B} & \multicolumn{1}{l}{C} & \multicolumn{1}{l}{D/E} \\
\midrule
Dataset I & 180,126 & 80,411 & 74,805 & 1 & 153,060 & 57,822 & 89,842 & 12 \\
Dataset II & 34,011 & 15,497 & 15,001 & 0 & 32,881 & 16,678 & 11,189 & 4,990 \\
Dataset III & 196,481 & 105,932 & 69,108 & 3 & 46,830 & 37,677 & 32,706 & 13 \\
\bottomrule
\end{tabular}
\end{table*}

\begin{table*}[htbp]
\centering
\caption{Top 5 ASN Ranges Distribution}
\label{tab:asn_dist}
\resizebox{\textwidth}{!}{
\begin{tabular}{@{}ll*{5}{l}@{}}
\toprule
\multirow{2}{*}{Dataset} & \multirow{2}{*}{Type} & \multicolumn{5}{c}{Top 5 ASN Ranges (IP Count)} \\
\cmidrule(lr){3-7}
 & & 1st & 2nd & 3rd & 4th & 5th \\
\midrule
\multirow{2}{*}{Dataset I} 
 & Benign & AS5 (41,672) & AS12 (38,840) & AS13 (35,965) & AS2 (32,601) & AS4 (30,728) \\
 & Malicious & AS12 (73,840) & AS4 (35,235) & AS6 (29,115) & AS2 (27,075) & AS5 (24,747) \\
\addlinespace
\multirow{2}{*}{Dataset II} 
 & Benign & AS12 (7,819) & AS13 (7,187) & AS5 (7,077) & AS2 (6,492) & AS4 (5,994) \\
 & Malicious & AS12 (7,662) & AS10 (6,371) & AS6 (5,784) & AS2 (4,856) & AS4 (4,671) \\
\addlinespace
\multirow{2}{*}{Dataset III} 
 & Benign & AS2 (48,126) & AS12 (40,053) & AS5 (37,565) & AS8 (32,219) & AS11 (31,514) \\
 & Malicious & AS12 (28,140) & AS10 (14,853) & AS2 (10,921) & AS4 (9,719) & AS6 (9,641) \\
\bottomrule
\end{tabular}
}
\end{table*}

\section{Methodology}
\begin{figure*}[!t]
\centering
\includegraphics[width=0.95\linewidth]{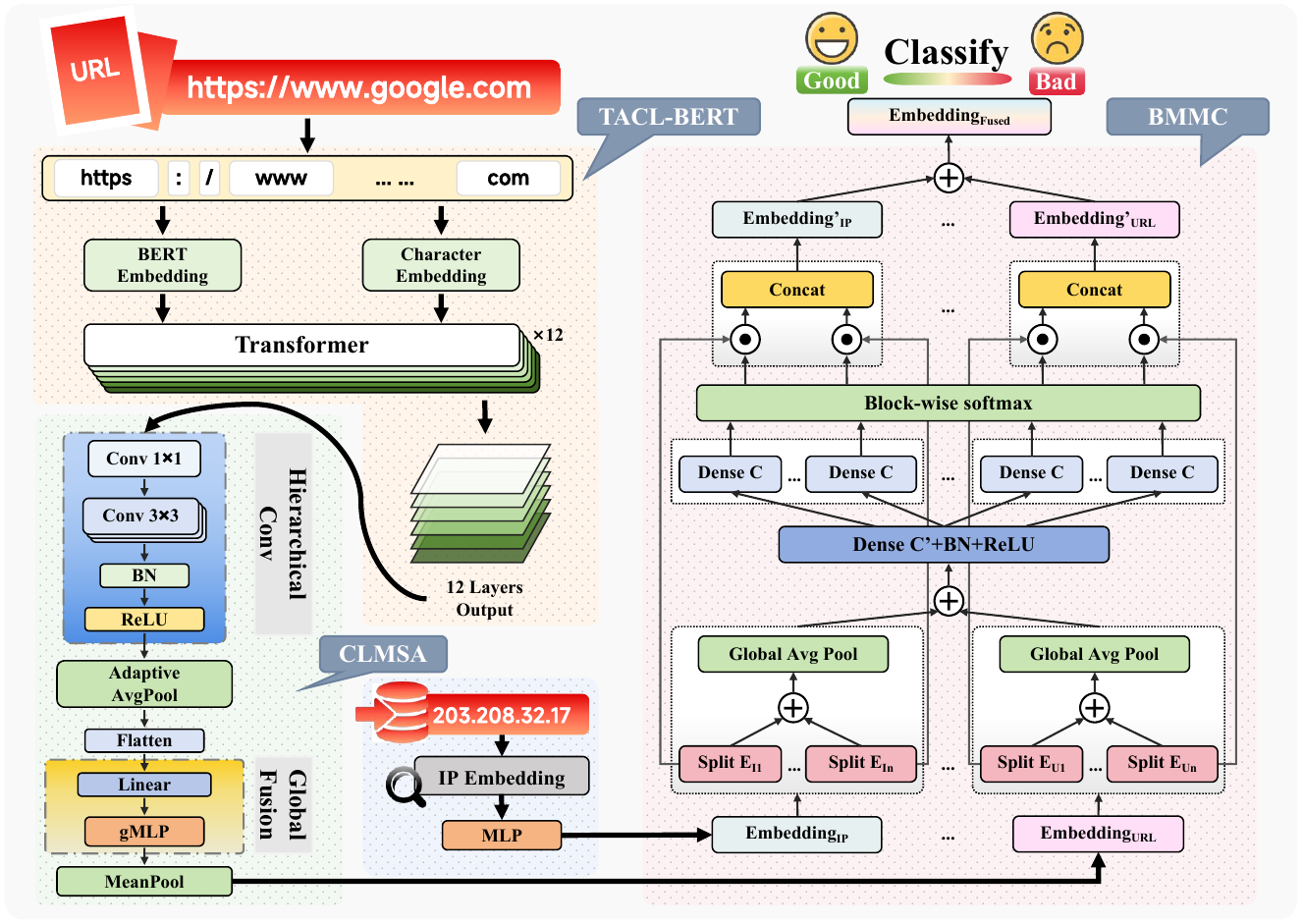}
\caption{Architecture: Composed of Three Core Modules.
TACL-BERT serves as the backbone encoder for token-aware subword representations; CLMSA hierarchically aggregates multi-layer features through convolution and gated MLPs; BMMC performs structured multi-scale attention fusion across URL and IP modalities.}
\label{fig_framework}
\end{figure*}

We present \textbf{CURL-IP}, an IP-enhanced multimodal malicious URL detection framework, whose architecture is illustrated in \Cref{fig_framework}. The proposed system comprises three core components: (1) a character-aware \textbf{TACL-BERT} encoder, (2) a hierarchical \textbf{CLMSA} feature aggregator, and (3) a multimodal \textbf{BMMC} fusion module.

\begin{itemize}
\item \textbf{TACL-BERT}: Our backbone encoder employs token-aware contrastive distillation to enhance subword representation quality, significantly improving discriminability and isotropy for robust encoding of adversarial URLs with token-level perturbations.

\item \textbf{CLMSA}: This feature aggregator combines hierarchical convolutional operations with gMLP blocks to effectively fuse multi-layer Transformer features, simultaneously capturing local character patterns and global contextual relationships.

\item \textbf{BMMC}: The cross-modal fusion module utilizes structured block-wise attention mechanisms to align URL and IP representations.
\end{itemize}

\subsection{Backbone Network: TACL-BERT}
TACL-BERT is a token-aware contrastive learning-enhanced BERT encoder, motivated by contrastive learning~\cite{r39}, that enables robust URL sequence representations. The model extends standard masked language modeling (MLM) through a novel contrastive token-level objective that simultaneously enhances the discriminativeness and isotropy of token embeddings - crucial properties for malicious URL detection where minor lexical variations can indicate fundamentally different semantic intents.

TACL-BERT adopts a student-teacher framework where the teacher encoder $f_{\text{T}}$ processes the original unmasked input sequence $x = [t_1, t_2, \ldots, t_m]$ to generate contextual token representations:
\begin{equation}
    h_i = f_{\text{T}}(x)_i \in \mathbb{R}^d
\end{equation}

The student encoder $f_{\text{S}}$ processes the identical input sequence with randomly masked tokens (following standard BERT masking protocol, e.g., 15\% replacement) and generates corresponding representations:
\begin{equation}
    \tilde{h}_i = f_{\text{S}}(\tilde{x})_i
\end{equation}

For each masked token $t_i$, the model aligns the student's representation $\tilde{h}_i$ with the corresponding teacher output $h_i$ while contrasting it against other sequence tokens. The token-aware contrastive loss formalizes this objective as:
\begin{equation}
\mathcal{L}_{\text{TaCL}} = - \sum_{i \in \mathcal{M}} \log \frac{\exp(\text{sim}(\tilde{h}_i, h_i)/\tau)}{\sum\limits_{j=1}^{m} \exp(\text{sim}(\tilde{h}_i, h_j)/\tau)}
\end{equation}
where:
\begin{itemize}
  \item $\mathcal{M}$ is the set of masked positions,
  \item $\text{sim}(u,v) = \frac{u^\top v}{\|u\|\|v\|}$ is the cosine similarity,
  \item $\tau$ is a temperature hyperparameter.
\end{itemize}

This loss encourages each masked token representation to be close to its own clean counterpart while being distant from other tokens within the same batch. It complements the standard MLM loss:
\begin{equation}
\mathcal{L}_{\text{MLM}} = - \sum_{i \in \mathcal{M}} \log P(t_i | \tilde{x})
\end{equation}

The final pretraining objective combines both terms:
\begin{equation}
\mathcal{L}_{\text{total}} = \mathcal{L}_{\text{MLM}} + \lambda \cdot \mathcal{L}_{\text{TaCL}}
\end{equation}
where $\lambda$ is a weighting coefficient controlling the contribution of the contrastive loss.

Through optimization of $\mathcal{L}_{\text{TaCL}}$, the model generates token embeddings with enhanced semantic discriminability and improved spatial distribution characteristics. This proves particularly valuable for malicious URL detection scenarios, where adversaries frequently employ subtle character manipulations (e.g., paypa1.com) to bypass security measures.

\subsection{CLMSA Module}
To enhance and fuse deep semantic features across different layers of the TACL-BERT encoder, we propose CLMSA consisting of two core components: (1) a hierarchical convolutional block that progressively extracts abstract local semantic representations while capturing inter-layer variations, and (2) a global-aware fusion block that effectively integrates cross-scale information. Unlike the standard TACL-BERT framework that only utilizes final-layer token embeddings, our extended architecture aggregates hidden representations from all intermediate layers of the student model. For a given URL input, we extract:

\begin{equation}
    \mathcal{X}_{\text{BERT}} = \texttt{Stack}(H^{(1)}, H^{(2)}, \ldots, H^{(12)}) \in \mathbb{R}^{B \times L \times T \times D}
\end{equation}

The hidden state $H^{(l)} \in \mathbb{R}^{B \times T \times D}$ represents the output from the $l$-th Transformer layer, where $B$ indicates batch size, $T$ specifies token sequence length, $L=12$ determines the total layer count, and $D=768$ defines the hidden dimension size. This multi-layer tensor forms a hierarchical semantic structure preserving both low-level and high-level contextual features. 

Prior to convolutional processing, we reorganize the tensor by rearranging its semantic and temporal dimensions through axis permutation:

\begin{equation}
    \mathcal{X}_0 = \texttt{Permute}(\mathcal{X}_{\text{BERT}}) \in \mathbb{R}^{B \times L \times D \times T}
\end{equation}

The tensor $\mathcal{X}_0$ is processed by a stack of four convolutional blocks. Each block includes a $3{\times}3$ convolution layer, followed by batch normalization and ReLU activation. The output channel size decreases progressively from 64 to 8, allowing hierarchical abstraction:

\begin{equation}
    \mathcal{X}_{i} = \sigma(\texttt{BN}_i(\texttt{Conv}_i(\mathcal{X}_{i-1}))), \quad i = 1,2,3,4
\end{equation}

where $\sigma(\cdot)$ denotes the ReLU activation function, and the transformation follows the channel reduction: $64 \rightarrow 32 \rightarrow 16 \rightarrow 8$.

To unify spatial dimensions for downstream processing, we apply adaptive average pooling to normalize feature maps:

\begin{equation}
    \mathcal{X}_{\text{pool}} = \texttt{AdaptiveAvgPool2D}(\mathcal{X}_4) \in \mathbb{R}^{B \times 8 \times 25 \times 96}
\end{equation}

The resulting tensor is reshaped to form a token-wise representation:

\begin{equation}
    \mathcal{X}_{\text{flat}} = \texttt{Reshape}(\mathcal{X}_{\text{pool}}) \in \mathbb{R}^{B \times 25 \times 768}
\end{equation}

We then apply a linear projection to reduce the feature dimension:

\begin{equation}
    \mathcal{X}_{\text{proj}} = \mathcal{X}_{\text{flat}} \cdot \mathbf{W}_p + \mathbf{b}_p \in \mathbb{R}^{B \times 25 \times 128}
\end{equation}

where $\mathbf{W}_p \in \mathbb{R}^{768 \times 128}$ is the projection matrix and $\mathbf{b}_p$ is the bias term. This produces a compact token sequence of length 25, each with 128-dimensional features.

To model long-range token dependencies and capture global contextual patterns, the projected sequence is passed through a lightweight gMLP layer:

\begin{equation}
    \mathcal{X}_{\text{gmlp}} = \texttt{gMLP}(\mathcal{X}_{\text{proj}}) \in \mathbb{R}^{B \times 25 \times 128}
\end{equation}

Finally, temporal average pooling is applied across the token axis to obtain a unified representation:

\begin{equation}
    f_{\text{url}} = \frac{1}{25} \sum_{t=1}^{25} \mathcal{X}_{\text{gmlp}}^{(t)} \in \mathbb{R}^{B \times 128}
\end{equation}

The output vector $f_{\text{url}}$ captures both local structural cues and global sequence-level semantics, and is passed to the downstream multimodal fusion module.

\subsection{IP Embedding}
The IP processing branch employs a lightweight MLP network operating in parallel with the URL encoder, transforming raw IP embeddings (sourced externally) into a latent representation matching the target dimension:
\begin{equation}
    f_{ip} = \text{ReLU}(W_{ip} \cdot \text{IP\_embed}) \in \mathbb{R}^{B \times 128}
\end{equation}
This parallel architecture enables simultaneous capture of network-level characteristics through IP analysis while maintaining content-based URL examination, with the ReLU-activated projection ensuring compatible feature spaces for subsequent multimodal fusion.

The design provides three key advantages: (1) preserving computational efficiency through lightweight MLP implementation, (2) maintaining dimensional consistency ($\mathbb{R}^{B \times 128}$) for downstream processing, and (3) enabling complementary threat detection through concurrent analysis of URL content and network origin attributes. The IP branch specifically enhances the detection of malicious patterns that may exhibit normal URL characteristics but originate from suspicious network infrastructures.

\subsection{BMMC Module}

We introduce BMMC, a novel fusion module that effectively combines URL semantic features with auxiliary IP-based characteristics through advanced cross-modal attention. Unlike conventional concatenation or early fusion methods, BMMC operates by decomposing multi-modal inputs into localized block-level units, where cross-modal attention weights at the block level are computed to selectively amplify or attenuate feature regions. This block-wise attention mechanism enables robust capture of local cross-modal correlations while mitigating overfitting to noisy modality components.

The module processes $M$ input modalities ${x^{(1)}, x^{(2)}, \ldots, x^{(M)}}$, where each $x^{(m)} \in \mathbb{R}^{B \times C_m \times D}$ represents the $m$-th modality with channel dimension $C_m$ and temporal/spatial extent $D$. Each modality is partitioned into fixed-size channel blocks $C_b$, with padding applied when necessary to ensure dimensional compatibility, yielding reshaped tensors $\tilde{x}^{(m)} \in \mathbb{R}^{B \times N_m \times C_b \times D}$ where $N_m = \lceil C_m / C_b \rceil$ denotes the resulting number of blocks per modality.

The BMMC then computes a global representation by summing the blocks of each modality and applying adaptive average pooling. The pooled summaries across all modalities are summed and transformed via a shared joint feature extractor consisting of a linear layer, batch normalization, and ReLU activation. This results in a compact global context vector $g$.
\begin{equation}
    g^{(m)} = \texttt{GAP}\left( \sum_{i=1}^{N_m} \tilde{x}_i^{(m)} \right), \quad g = \sum_{m=1}^M g^{(m)}
\end{equation}

For each block, a learned projection transforms the global representation $g$ into a channel-wise attention vector $\alpha = [\alpha_1, \dots, \alpha_M]$ using a dedicated linear layer, where $M$ is the number of feature blocks. These raw attention scores are normalized across blocks using a softmax function to ensure comparability. To avoid overly suppressing any block, the attention weights are further scaled to the range $[\alpha_{\min}, 1]$ via the following formula:

\begin{equation}
\alpha_i = \alpha_{\min} + (1 - \alpha_{\min}) \cdot \alpha_i
\end{equation}

Here, $\alpha_i$ denotes the attention weight for the $i$-th block after scaling, and $\alpha_{\min} \in [0, 1]$ is a hyperparameter controlling the minimum attention each block can receive.

During training, a block-level dropout is applied to encourage robustness. A binary mask $m_i$ is sampled from a Bernoulli distribution with keep probability $1 - p$, where $p$ is the dropout rate. This mask randomly disables certain blocks by zeroing them out. Finally, the attention-weighted and optionally masked blocks are reshaped back to their original form and passed to the next stage.

\begin{equation}
    \tilde{x}^{(m)}_i \leftarrow \tilde{x}^{(m)}_i \cdot m_i \cdot \alpha_i, \quad m_i \sim \texttt{Bernoulli}(1 - p)
\end{equation}
\begin{equation}
    \hat{x}^{(m)} = \texttt{Reshape}(\tilde{x}^{(m)}), \quad \hat{x}^{(m)} \in \mathbb{R}^{B \times C_m \times D}
\end{equation}

This architecture empowers BMMC to dynamically adjust feature region weights according to global contextual information while selectively suppressing non-informative components through learned dropout mechanisms. The modular block-based design ensures scalability to inputs of varying dimensions and maintains robustness against heterogeneous statistical distributions across different input modalities. The complete BMMC framework is visually presented in \Cref{fig_framework}.

\section{Experiments}
This section presents comprehensive experimental evaluations of our proposed method, including comparative analyses with baseline approaches. The investigation encompasses four key aspects: (1) data scale dependency analysis across varying training set sizes, (2) multi-class classification performance, (3) robustness assessment against adversarial examples, and (4) utility evaluation for recently active malicious URLs. All experiments were conducted under controlled conditions to ensure reproducible results.

\subsection{Setup}
\subsubsection{Experimental Infrastructure}
The TACL-BERT model was pre-trained on the English Wikipedia corpus (12GB, ~2.5B words), with continuous optimization of the student model during this phase. Fine-tuning employed the following hyperparameters: batch size 16, AdamW optimizer (initial learning rate 2e-5, weight decay 1e-4), dropout rate 0.1, and 10 training epochs. All experiments were conducted using PyTorch 2.0 with CUDA 11.8 acceleration on NVIDIA RTX 3090 GPUs, implemented in Python 3.8. Model selection was performed based on optimal validation loss performance.

\subsubsection{Baseline Methods}
We evaluate our proposed approach against four state-of-the-art baselines: URLNet, TransURL, URLBERT, and TextCNN. For fair comparison, we implemented all baseline models using their original GitHub repositories without architectural modifications or hyperparameter tuning. Consistent with the original URLNet study, we employed embedding mode 5 (the most comprehensive configuration) due to its demonstrated superior performance. All models were trained and evaluated on identical datasets under consistent experimental conditions.

\subsubsection{Evaluation Metrics}
We adopt six standard metrics for evaluating malicious URL detection performance: accuracy, precision, recall, F1 score, ROC curve, and AUC. These metrics are formally defined as follows:

\begin{equation}\label{eq:accuracy}
\text{Accuracy} = \frac{\text{TP} + \text{TN}}{\text{TP} + \text{FP} + \text{TN} + \text{FN}}
\end{equation}

\begin{equation}\label{eq:precision}
\text{Precision} = \frac{\text{TP}}{\text{TP} + \text{FP}}
\end{equation}

\begin{equation}\label{eq:recall}
\text{Recall} = \frac{\text{TP}}{\text{TP} + \text{FN}}
\end{equation}

\begin{equation}\label{eq:f1}
\text{F1 Score} = 2 \times \frac{\text{Precision} \times \text{Recall}}{\text{Precision} + \text{Recall}}
\end{equation}
The ROC curve plots the true positive rate (TPR) against false positive rate (FPR):
\begin{equation}\label{eq:fpr}
\text{FPR} = \frac{\text{FP}}{\text{FP} + \text{TN}}
\end{equation}
\begin{equation}\label{eq:tpr}
\text{TPR} = \frac{\text{TP}}{\text{TP} + \text{FN}}
\end{equation}
AUC (area under curve) values range from 0.5 (random guessing) to 1.0 (perfect discrimination).

Where:
\begin{itemize}
\item TP: Malicious URLs correctly detected
\item FP: Benign URLs falsely flagged as malicious
\item TN: Benign URLs correctly identified
\item FN: Malicious URLs missed by detection
\end{itemize}

\subsection{Comparison with Baselines}
We present a comprehensive comparison of our method against baseline approaches for both binary and multi-class malicious URL detection scenarios.

\subsubsection{Binary Classification}
\begin{figure*}[!t]
    \centering

    \subfloat[Accuracy]{%
        \includegraphics[width=0.32\textwidth]{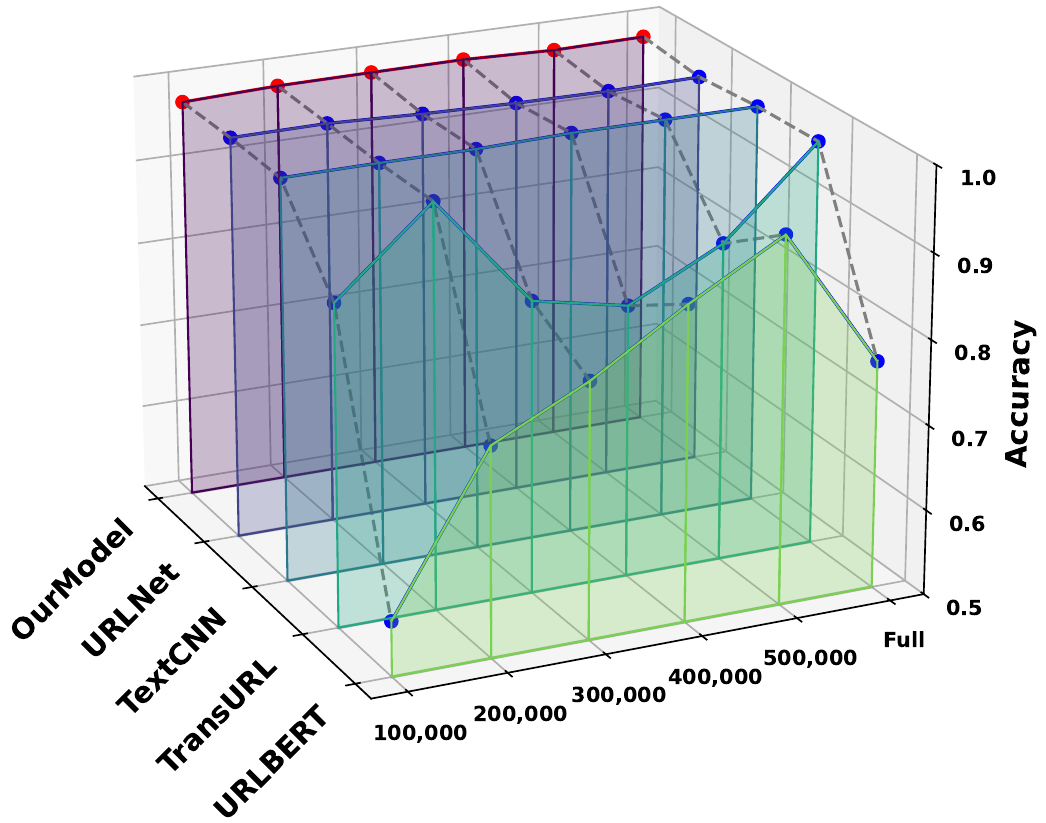}%
        \label{fig:acc}
    } \hfill
    \subfloat[Precision]{%
        \includegraphics[width=0.32\textwidth]{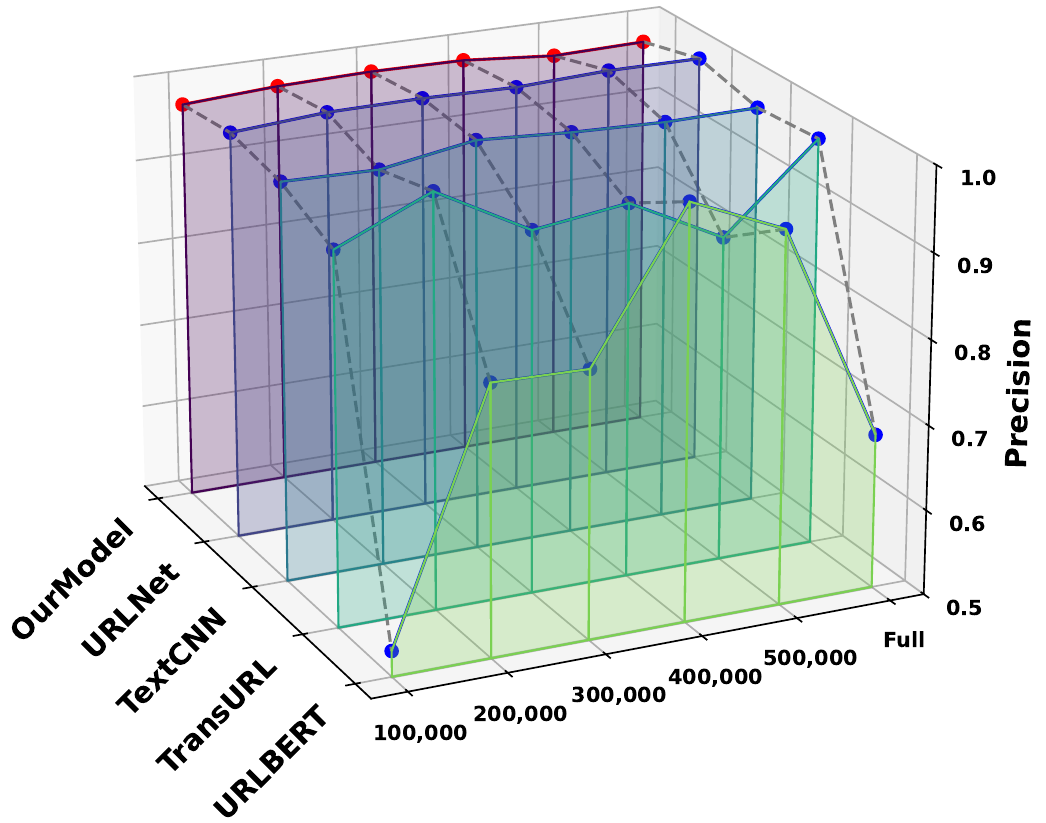}%
        \label{fig:prec}
    } \hfill
    \subfloat[Recall]{%
        \includegraphics[width=0.32\textwidth]{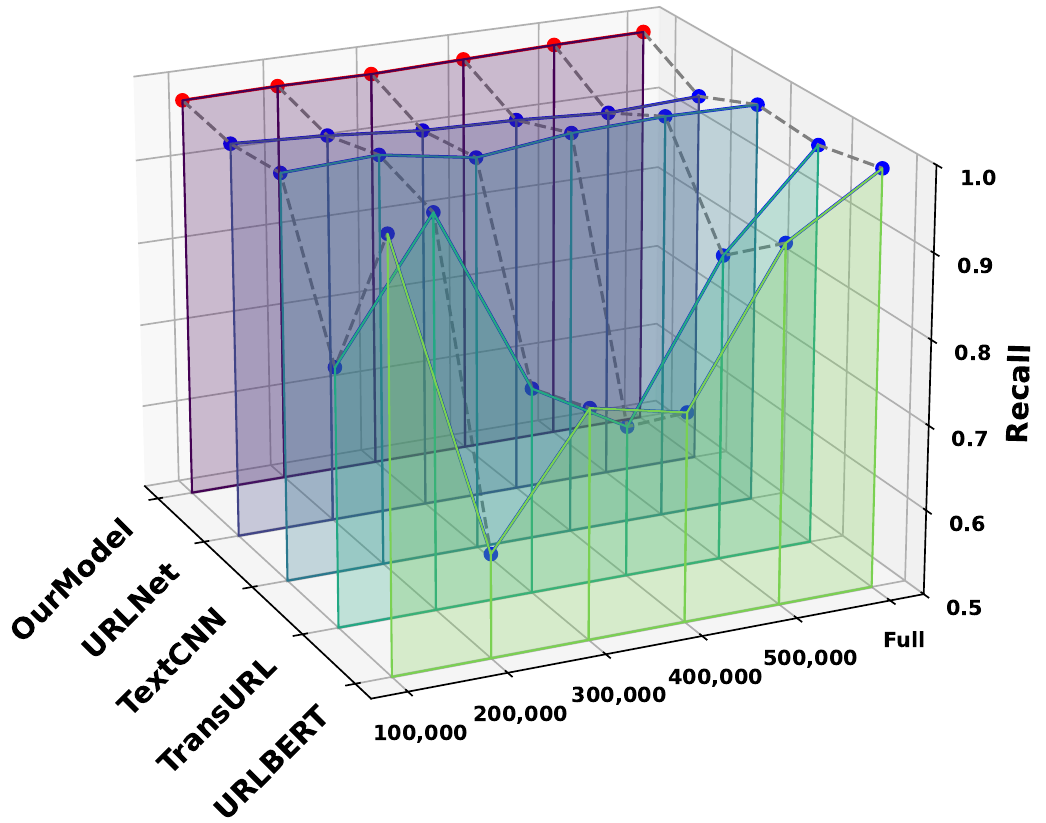}%
        \label{fig:rec}
    }

    \vspace{1em}

    \subfloat[F1 Score]{%
        \includegraphics[width=0.32\textwidth]{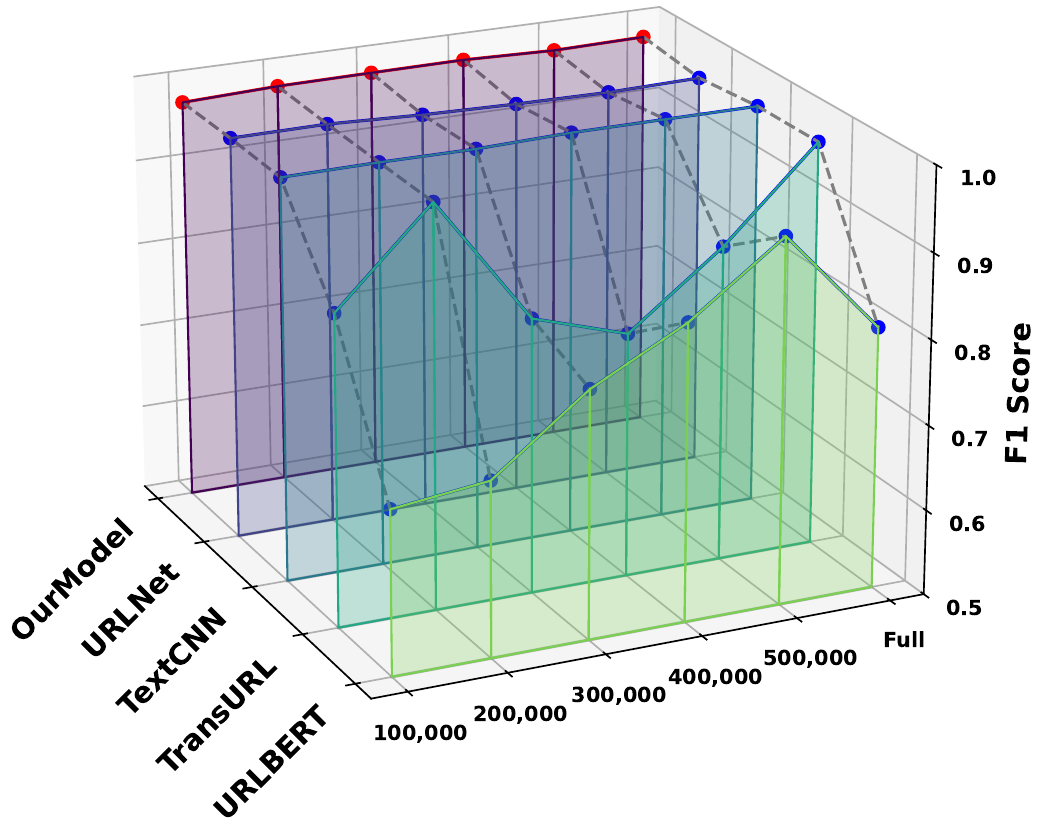}%
        \label{fig:f1}
    } \hspace{0.1\textwidth}
    \subfloat[AUC]{%
        \includegraphics[width=0.32\textwidth]{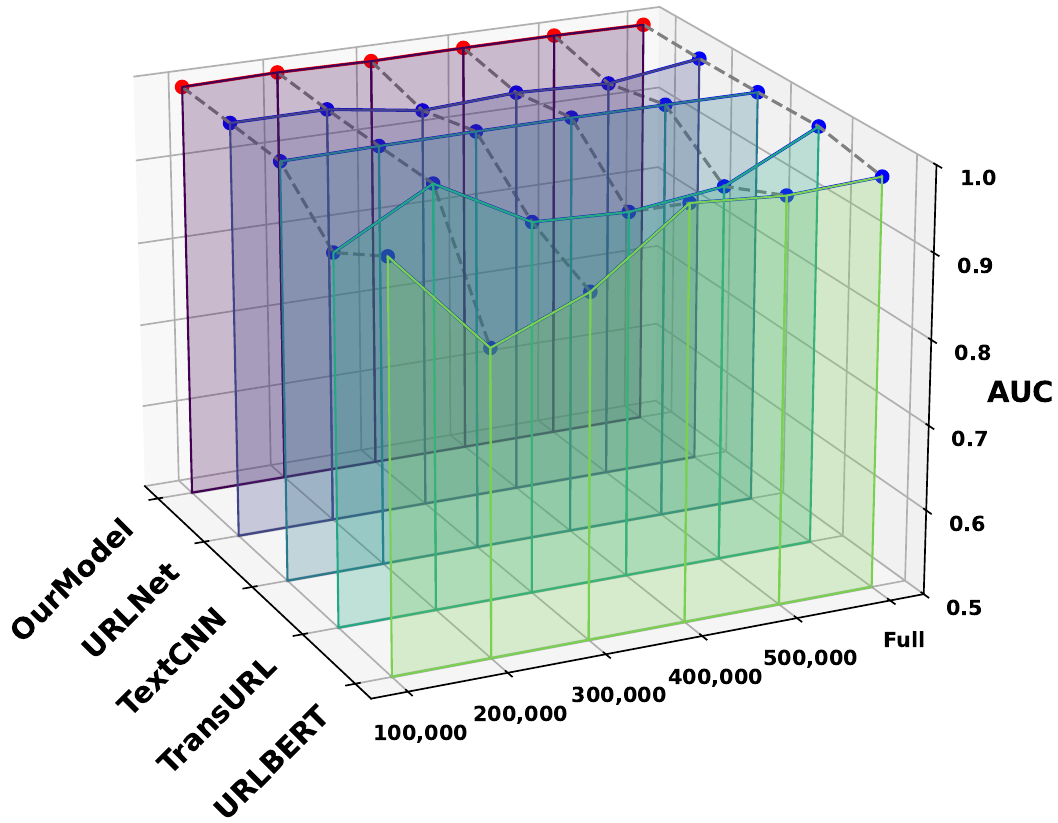}%
        \label{fig:auc}
    }

    \caption{Model performance comparison across different training dataset sizes.}
    \label{fig:model-performance-comparison}
\end{figure*}

To examine the impact of training data volume, we conduct scaled experiments with progressively increasing dataset sizes from 100,000 samples to 500,000 samples in 100,000-sample increments. Each configuration maintains identical test conditions, with all models evaluated on a fixed test set to ensure consistent comparison. As evidenced in \Cref{tab:size} and \Cref{fig:model-performance-comparison}, this systematic evaluation reveals the performance scaling characteristics relative to training data quantity.

As shown in \Cref{tab:size}, our method consistently outperforms all baseline models across all training sizes. Even with only 100,000 training samples, our model achieves an F1-score of 0.9758 and an AUC of 0.9941, significantly higher than TransURL (F1-score 0.8664, AUC 0.9302) and URLBERT (F1-score 0.6926, AUC 0.9743). As the training data increases, our model continues to exhibit strong and stable performance. At 200,000 and 300,000, it achieves F1-scores of 0.9798 and 0.9807, with corresponding AUCs of 0.9962 and 0.9947, both of which are the highest among all models. Notably, our model maintains its superiority even at the 500,000 training size, reaching an F1-score of 0.9780 and an AUC of 0.9960.In particular, as shown in \Cref{fig:model-performance-comparison}, our model exhibits extremely strong recall (0.9783-0.9858) on all data scales.

When trained on the full dataset, our model attains the best overall results: an accuracy of 0.9799, precision of 0.9737, recall of 0.9858, F1-score of 0.9797, and AUC of 0.9946. These results indicate that our model not only benefits from increased training data, but also maintains excellent generalization. In contrast, URLBERT demonstrates poor robustness, with highly unstable performance, particularly under limited data settings (e.g., AUC of 0.9743 at 100,000 vs. 0.9749 at 500,000). Our model's consistent performance across all scales highlights its ability to effectively capture discriminative patterns from both large and small datasets, making it a strong choice for real-world binary malicious URL detection tasks.

To further evaluate the detection effectiveness of our model under different false positive rates (FPRs), we report the true positive rate (TPR) at various FPR levels in \Cref{tab:performance}. This metric is particularly relevant in high-stakes security applications, where maintaining low false alarm rates is critical.

We observe that while baselines such as URLNet and TextCNN achieve competitive accuracy and F1-score on larger datasets, their performance under strict false positive constraints or limited data conditions is notably less stable. As shown in \Cref{tab:size}, although TextCNN achieves an F1-score of 0.9733 at 100,000 and 0.9769 on the full dataset, the gain becomes marginal as training size increases. Across all data scales, our method demonstrates clear superiority, especially under low-FPR conditions. For instance, at an extremely low FPR of 0.001, our model achieves a TPR of 0.4958 at 100,000, which is over 35 times higher than TransURL (0.0138) and substantially higher than URLBERT (0.1364) and TextCNN (0.0810). As the training size increases to 300,000, the TPR@0.001 of our model rises sharply to 0.8677, while other baselines remain significantly behind.

At FPR = 0.0001, our model still maintains a detectable signal even under severe false-positive constraints, reaching a TPR of 0.5478 at 300,000 and 0.9263 on the full dataset, while all baselines struggle in this regime. For example, at full scale, TransURL and TextCNN only achieve TPRs of 0.1227 and 0.0487 respectively, confirming that our model is particularly well-suited for low-FPR security scenarios.

These results highlight that beyond achieving high average performance, our model is also robust and highly effective under practical deployment constraints, where very low false positive rates are required. The ability to maintain high detection rates under strict FPR thresholds further validates the reliability of our approach in real-world malicious URL detection tasks.These superior results, especially under low-resource and low-FPR settings, can be attributed to three key components of our model. First, the hierarchical encoding from TACL-BERT captures rich contextual dependencies across all Transformer layers, enabling the model to generalize well from limited examples. Second, the CLMSA module fuses multi-scale semantics through deep convolution and gated MLP operations, allowing both local features and global structures to be captured effectively. Third, the BMMC module facilitates fine-grained, block-level cross-modal attention fusion, dynamically selecting the most informative subregions from both URL and IP modalities. Together, these mechanisms enable our model to maintain high sensitivity and generalization, even with limited supervision and under deployment-level detection constraints.

\begin{table}[htbp]
\centering
\caption{Performance Comparison of Different Methods}
\label{tab:size}
\resizebox{\columnwidth}{!}{%
\begin{tabular}{l l c c c c c}
\toprule
Training Size & Method & Accuracy & Precision & Recall & F1-score & AUC \\
\midrule
\multirow{5}{*}{100,000} & Our & \textbf{0.9762} & 0.9733 & 0.9783 & \textbf{0.9758} & \textbf{0.9941} \\
 & TransURL & 0.8762 & 0.9360 & 0.8029 & 0.8644 & 0.9328 \\
 & URLBERT & 0.5643 & 0.5300 & 0.9990 & 0.6926 & 0.9743 \\
 & URLNet & 0.9761 & 0.9820 & 0.9692 & 0.9756 & 0.9932 \\
 & TextCNN & 0.9727 & 0.9684 & 0.9783 & 0.9733 & 0.9916 \\
\midrule
\multirow{5}{*}{200,000} & Our & \textbf{0.9802} & 0.9798 & 0.9799 & \textbf{0.9798} & \textbf{0.9962} \\
 & TransURL & 0.9737 & 0.9849 & 0.9612 & 0.9729 & 0.9936 \\
 & URLBERT & 0.7453 & 0.8165 & 0.6211 & 0.7055 & 0.8554 \\
 & URLNet & 0.9771 & 0.9901 & 0.9629 & 0.9763 & 0.9935 \\
 & TextCNN & 0.9737 & 0.9660 & 0.9829 & 0.9744 & 0.9931 \\
\midrule
\multirow{5}{*}{300,000} & Our & \textbf{0.9811} & 0.9822 & \textbf{0.9792} & \textbf{0.9807} & \textbf{0.9947} \\
 & TransURL & 0.8432 & 0.9244 & 0.7415 & 0.8229 & 0.9336 \\
 & URLBERT & 0.7999 & 0.8135 & 0.7690 & 0.7906 & 0.9005 \\
 & URLNet & 0.9728 & 0.9911 & 0.9531 & 0.9717 & 0.9768 \\
 & TextCNN & 0.9737 & 0.9841 & 0.9639 & 0.9739 & 0.9941 \\
\midrule
\multirow{5}{*}{400,000} & Our & \textbf{0.9817} & 0.9808 & \textbf{0.9819} & \textbf{0.9813} & \textbf{0.9958} \\
 & TransURL & 0.8201 & 0.9387 & 0.6780 & 0.7874 & 0.9285 \\
 & URLBERT & 0.8691 & 0.9845 & 0.7452 & 0.8483 & 0.9830 \\
 & URLNet & 0.9701 & 0.9889 & 0.9499 & 0.9690 & 0.9825 \\
 & TextCNN & 0.9766 & 0.9774 & 0.9765 & 0.9770 & 0.9944 \\
\midrule
\multirow{5}{*}{500,000} & Our & \textbf{0.9783} & 0.9715 & \textbf{0.9846} & \textbf{0.9780} & \textbf{0.9960} \\
 & TransURL & 0.8754 & 0.8823 & 0.8612 & 0.8716 & 0.9415 \\
 & URLBERT & 0.9308 & 0.9370 & 0.9212 & 0.9290 & 0.9749 \\
 & URLNet & 0.9690 & 0.9936 & 0.9430 & 0.9676 & 0.9782 \\
 & TextCNN & 0.9765 & 0.9737 & 0.9804 & 0.9770 & 0.9940 \\
\midrule
\multirow{5}{*}{Full} & Our & \textbf{0.9799} & 0.9737 & 0.9858 & \textbf{0.9797} & 0.9946 \\
 & TransURL & 0.9775 & 0.9807 & 0.9733 & 0.9770 & 0.9948 \\
 & URLBERT & 0.7675 & 0.6813 & 0.9894 & 0.8070 & 0.9800 \\
 & URLNet & 0.9709 & 0.9930 & 0.9475 & 0.9697 & 0.9930 \\
 & TextCNN & 0.9765 & 0.9751 & 0.9787 & 0.9769 & 0.9934 \\
\bottomrule
\end{tabular}%
}
\end{table}

\begin{table}[htbp]
\centering
\caption{Model Performance at Different Training Sizes (TPR @ FPR Levels)}
\label{tab:performance}
\begin{tabular}{@{}l l *{4}{S[table-format=1.4]} @{}}
\toprule
\multirow{2}{*}{Training Size} & \multirow{2}{*}{Method} & \multicolumn{4}{c}{TPR @ FPR Level} \\
\cmidrule(lr){3-6}
 & & {0.0001} & {0.001} & {0.01} & {0.1} \\
\midrule

\multirow{4}{*}{100,000} 
 & Our & 0.0162 & \textbf{0.4958 }& \textbf{0.9558} & 0.9900 \\
 & TransURL & 0.0013 & 0.0138 & 0.4737 & 0.8733 \\
 & URLBERT & 0.0866 & 0.1364 & 0.5360 & 0.9556 \\
 & TextCNN & 0.0394 & 0.0810 & 0.8189 & 0.9947 \\
\midrule

\multirow{4}{*}{200,000} 
 & Our & 0.0143 & \textbf{0.7756} & \textbf{0.9670} & 0.9923 \\
 & TransURL & 0.0639 & 0.4947 & 0.9432 & 0.9893 \\
 & URLBERT & 0.1019 & 0.1588 & 0.4290 & 0.5933 \\
 & TextCNN & 0.0685 & 0.2016 & 0.8850 & 0.9945 \\
\midrule

\multirow{4}{*}{300,000} 
 & Our & \textbf{0.5478} & \textbf{0.8677} & \textbf{0.9721} & 0.9903 \\
 & TransURL & 0.0006 & 0.0181 & 0.5219 & 0.8321 \\
 & URLBERT & 0.1253 & 0.1961 & 0.3905 & 0.6738 \\
 & TextCNN & 0.0459 & 0.2263 & 0.9253 & 0.9942 \\
\midrule

\multirow{4}{*}{400,000} 
 & Our & \textbf{0.4708} & \textbf{0.8460} & \textbf{0.9729} & 0.9915 \\
 & TransURL & 0.0042 & 0.1596 & 0.4989 & 0.7936 \\
 & URLBERT & 0.0079 & 0.1382 & 0.6695 & 0.9741 \\
 & TextCNN & 0.0988 & 0.2755 & 0.9203 & 0.9950 \\
\midrule

\multirow{4}{*}{500,000} 
 & Our & \textbf{0.5635} & \textbf{0.8263} & \textbf{0.9742} & 0.9919 \\
 & TransURL & 0.0657 & 0.1985 & 0.5268 & 0.8489 \\
 & URLBERT & 0.1665 & 0.5811 & 0.8172 & 0.9385 \\
 & TextCNN & 0.1015 & 0.2446 & 0.9129 & 0.9954 \\
\midrule

\multirow{4}{*}{Full} 
 & Our & \textbf{0.9263} & \textbf{0.9263} & \textbf{0.9751} & 0.9928 \\
 & TransURL & 0.1227 & 0.7019 & 0.9541 & 0.9916 \\
 & URLBERT & 0.2085 & 0.5223 & 0.8514 & 0.9474 \\
 & TextCNN & 0.0487 & 0.1251 & 0.9130 & 0.9945 \\
\bottomrule
\end{tabular}
\end{table}

\subsubsection{Multi-class Classification}
To assess our method against sophisticated cyber threats, we performed multi-class classification using a dataset containing benign, malicious, and phishing URLs. The dataset exhibits significant class imbalance with 420,000 benign samples, 94,086 phishing samples, and 23,645 malicious samples, presenting substantial challenges for model generalization, particularly for the underrepresented malicious and phishing categories.

\begin{figure*}[!t]
    \centering

    \subfloat[Benign]{%
        \includegraphics[width=0.48\textwidth]{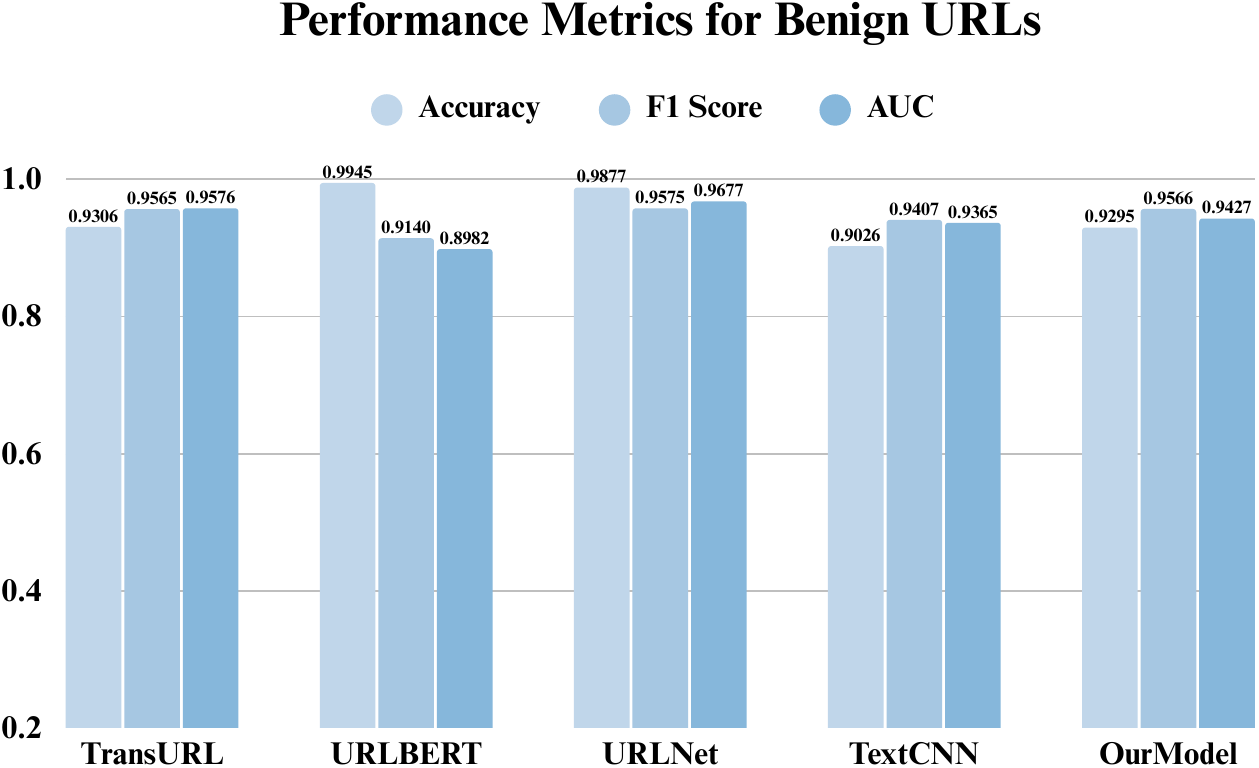}%
        \label{fig:benign}
    } \hfill
    \subfloat[Malicious]{%
        \includegraphics[width=0.48\textwidth]{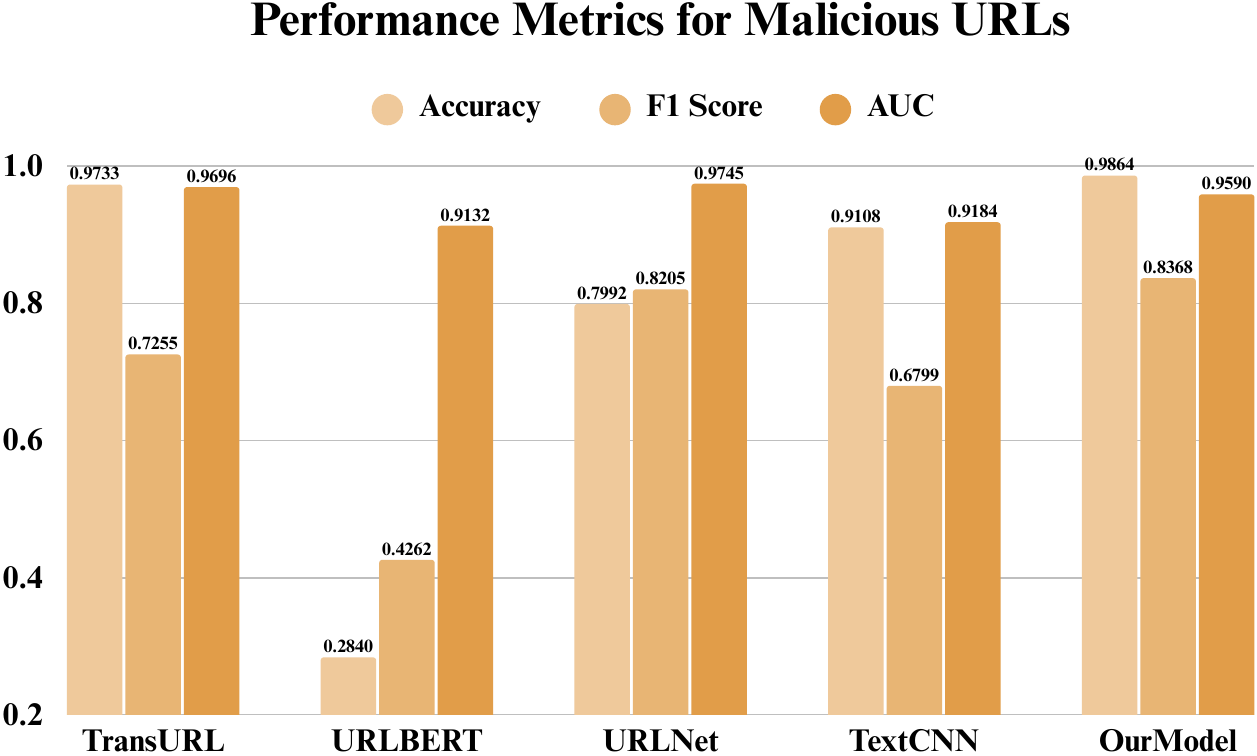}%
        \label{fig:malicious}
    } \hfill

    \vspace{1em}
    
    \subfloat[Phishing]{%
        \includegraphics[width=0.48\textwidth]{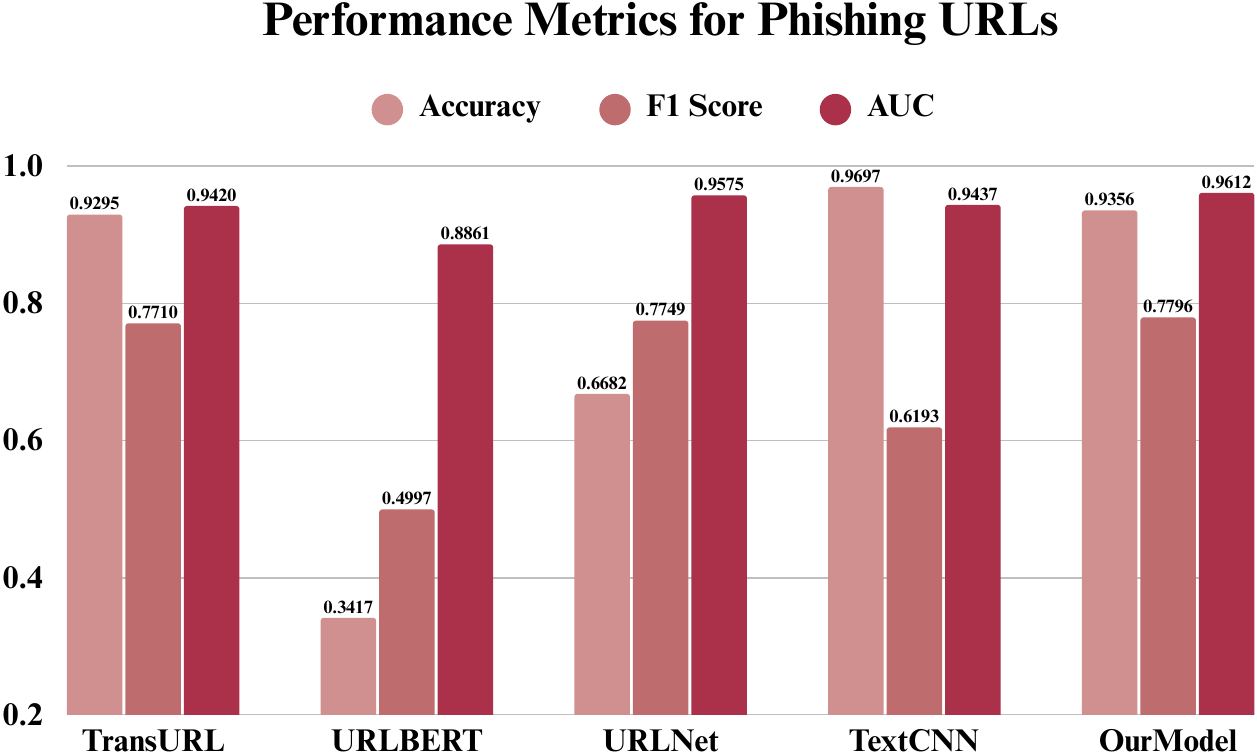}%
        \label{fig:phishing}
    }\hfill
    \subfloat[Macro ROC]{%
        \includegraphics[width=0.48\textwidth]{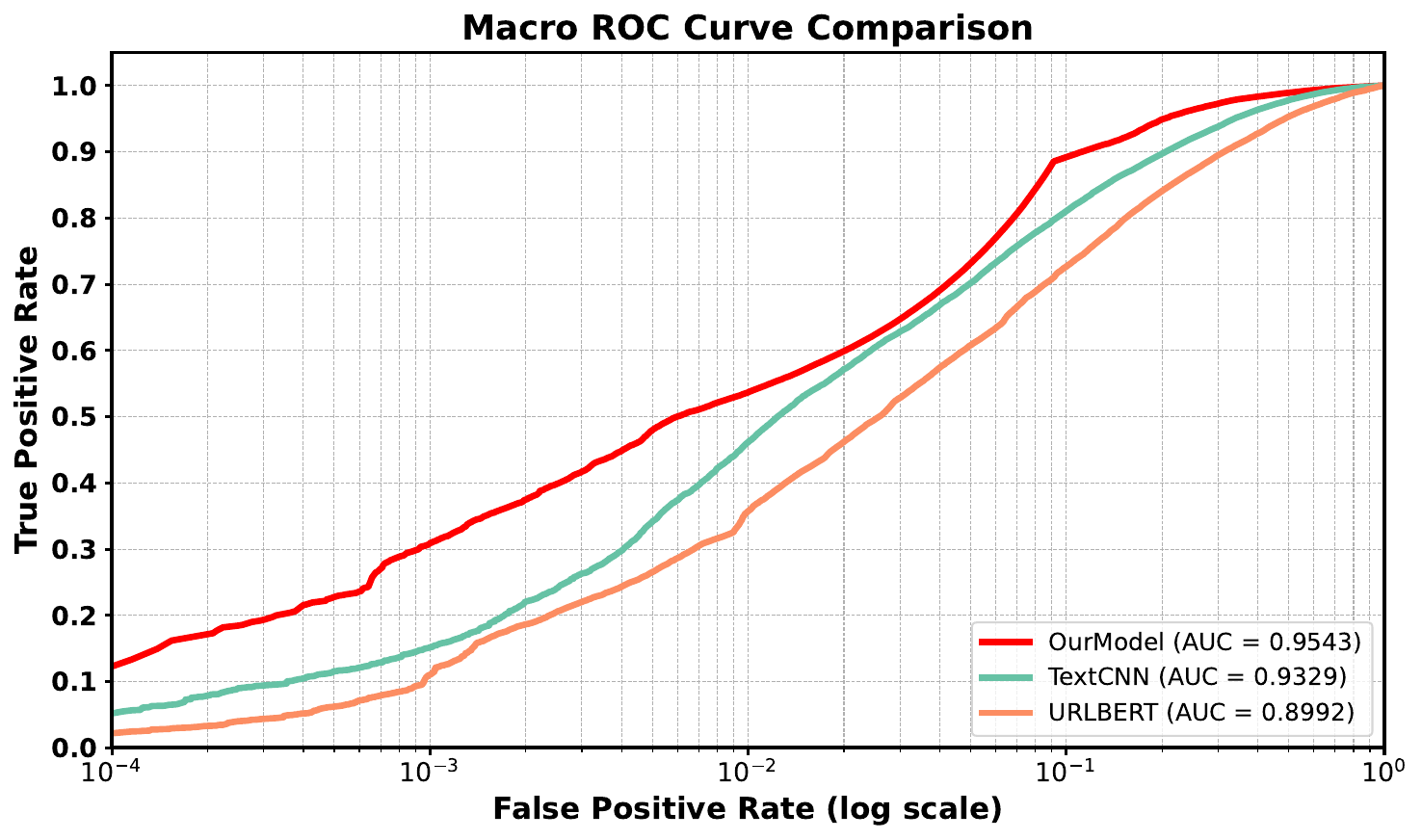}%
        \label{fig:macro_roc}
    }

    \caption{Detection results comparison of baseline methods and our proposed approach on multi-class dataset.}
    \label{fig:detection_results}
\end{figure*}

\Cref{fig:detection_results} compares the per-class performance of our model against four baselines (URLBERT, TextCNN, URLNet, and TransURL) using accuracy, F1-score, and AUC metrics. Our approach demonstrates consistent superiority across all categories, achieving 92.95\% accuracy and 95.66\% F1-score for benign URLs, 98.64\% accuracy and 83.68\% F1-score for malicious URLs, and 93.56\% accuracy and 77.96\% F1-score for phishing URLs. The model maintains robust discriminative ability with AUC values above 94.2\% for all classes, confirming its effectiveness despite the challenging class imbalance.

Compared to our method, TransURL shows relatively high AUC values (e.g., 95.76\%, 96.96\%, and 94.20\% across the three classes), but its F1-score on phishing URLs is considerably lower due to a sharp decline in classification accuracy. URLBERT exhibits a more severe performance gap: although it performs reasonably well on benign URLs (accuracy 99.45\%, AUC 89.82\%), its performance on malicious and phishing classes is poor, with accuracy below 35\% and F1-scores falling below 50\%. This suggests that URLBERT suffers from overfitting to the majority class and lacks robustness under class imbalance.

TextCNN performs better than URLBERT, with relatively balanced accuracy across classes, but its F1-score on phishing URLs remains low (around 62\%), reflecting limited discriminative power for minority threats. URLNet shows strong performance on benign and malicious URLs (F1-scores of 95.75\% and 82.05\%, respectively), yet its phishing F1-score is still suboptimal at 77.49\%, and below that of our method.

To further illustrate general classification capability across all classes, we compute and compare the macro-averaged ROC curves for our method and selected baselines, as shown in \Cref{fig:macro_roc}. Our model achieves the highest macro-AUC of 0.9543, significantly outperforming TextCNN (0.9329) and URLBERT (0.8992), confirming its superior ability to distinguish between URL classes even under strict evaluation settings.

Overall, our model demonstrates comprehensive detection capabilities, outperforming baselines across all threat categories while maintaining robust macro-level performance. These results validate its practical effectiveness for real-world malicious URL detection systems requiring reliable classification of diverse threat types.







\subsection{Robustness Evaluation Under Adversarial Conditions}
Adversaries frequently bypass security systems through carefully crafted input perturbations that preserve malicious functionality while evading detection. To rigorously evaluate our framework's defensive capabilities, we develop an adversarial benchmark based on the Compound Attack methodology \citep{r1}, incorporating strategic modifications from the GramBeddings approach \citep{r2} to simulate realistic evasion techniques encountered in operational cybersecurity environments. 

We implement an efficient processing pipeline that integrates a lightweight subword tokenizer with rule-based domain extraction (using \texttt{tld}) for precise URL component isolation. Adversarial samples are generated through strategic insertion of evasion characters (e.g., hyphens) between subword tokens within the domain segment, effectively simulating prevalent character-level obfuscation techniques while maintaining: (1) visual coherence of the modified domains, and (2) minimal structural deviation from legitimate patterns. To enhance evaluation realism, we derive pseudo-random IP addresses algorithmically from each adversarial URL's hash signature, ensuring consistent network context representation. The resulting evaluation set rigorously tests detector robustness against sophisticated evasion tactics.

As shown in \Cref{tab:adversarial_performance}, our model demonstrates superior robustness against adversarial attacks. While baseline models suffer significant performance degradation—URLBERT's accuracy drops to 68.53\% with an AUC of 85.29\%—our method maintains an accuracy of 93.95\% and an AUC of 98.12\%, outperforming URLBERT by over 24\% in accuracy. Although TransURL achieves a high recall of 96.58\%, its low precision (82.55\%) reflects a high false positive rate. In contrast, our model strikes a better balance between precision and recall, achieving the highest F1-score of 93.93\%. These results clearly demonstrate that our model achieves the best balance between detection sensitivity and reliability in adversarial settings, highlighting its strong generalization and resistance to evasion attacks through domain perturbations.

\begin{table}[htbp]
\centering
\caption{Performance under adversarial attack}
\label{tab:adversarial_performance}
\begin{tabular}{@{}l *{5}{S[table-format=1.4]} @{}}
\toprule
Method & {Accuracy} & {Precision} & {Recall} & {F1-score} & {AUC} \\
\midrule
Our & \textbf{0.9395} & \textbf{0.9427} & 0.9359 & \textbf{0.9393} & \textbf{0.9812} \\
TransURL & 0.8808 & 0.8255 & 0.9658 & 0.8902 & 0.9699 \\
URLBERT & 0.6853 & 0.6183 & 0.9685 & 0.7548 & 0.8529 \\
\bottomrule
\end{tabular}
\end{table}

\subsection{Practical Validation Case Study}
To further assess the practical utility of our model, we conducted a case study on 30 live phishing URLs recently reported and verified on PhishTank. These URLs reflect real-world threats and exhibit diverse structures, including IP-based domains, misleading subdomains, and unusual top-level domains such as \texttt{.shop}, \texttt{.space}, and \texttt{.cfd}. As summarized in Table~\ref{tab:case}, we identified 12 representative examples that were misclassified by at least one baseline model, demonstrating our method's improved detection capability against sophisticated real-world phishing attempts.

Among the baselines, URLNet failed to detect 4 malicious URLs, often misclassifying those with uncommon TLDs or shorter structures. This suggests URLNet may overly rely on surface-level heuristics such as URL length or common token patterns, leading to poor generalization when facing non-standard formats. Similarly, TransURL and URLBERT showed inconsistent performance, each misclassifying 4 or more URLs. Their errors include visually deceptive domains such as \texttt{allegrolokalnie.pl-oferta279529.cfd} and \texttt{centratipcreativess.pages.dev}, indicating difficulty in handling fine-grained obfuscation and rare domain compositions. TextCNN performed slightly better in terms of count but failed to detect several complex and IP-based threats, including \texttt{http://www.allegro.pl-conformation103.shop}.

In contrast, our proposed model successfully identified all 30 malicious URLs, demonstrating consistent robustness and adaptability across a variety of real-world phishing formats. This superior performance can be attributed to our comprehensive architecture, which includes character-aware token embeddings, hierarchical semantic modeling, and fine-grained cross-modal alignment. Together, these components enable our model to capture nuanced lexical cues, global semantic structures, and auxiliary network-level information, resulting in more reliable detection even for previously unseen or intentionally obfuscated threats.

\begin{table*}[t]
\caption{Case Study: Misclassified URLs Across Different Models}
\label{tab:case}
\centering
\begin{tabular}{lccccc}
\toprule
\textbf{Malicious url} & \textbf{TextCNN} & \textbf{TransURL} & \textbf{URLBERT} & \textbf{URLNet} & \textbf{Our} \\ 
\midrule
http://37.49.228.176/hiddenbin/boatnet.x86 & $\checkmark$ & $\times$ & $\checkmark$ & $\checkmark$ & $\checkmark$ \\ 
http://allegrolokalnie.pl-oferta279529.cfd & $\checkmark$ & $\checkmark$ & $\checkmark$ & $\times$ & $\checkmark$ \\ 
http://welltetarxor.webflow.io & $\checkmark$ & $\checkmark$ & $\times$ & $\times$ & $\checkmark$ \\ 
http://centratipcreativess.pages.dev & $\times$ & $\times$ & $\checkmark$ & $\times$ & $\checkmark$ \\ 
https://nzyyymwl.za.com/espera.html & $\checkmark$ & $\checkmark$ & $\checkmark$ & $\times$ & $\checkmark$ \\ 
http://www.allegro.pl-conformation103.shop & $\times$ & $\times$ & $\times$ & $\times$ & $\checkmark$ \\
http://coupangshope.shop & $\times$ & $\times$ & $\checkmark$ & $\times$ & $\checkmark$ \\ 
http://thevihshub.com/short/ & $\times$ & $\checkmark$ & $\checkmark$ & $\checkmark$ & $\checkmark$ \\ 
https://www.bitcapitalmine.com/ & $\checkmark$ & $\checkmark$ & $\checkmark$ & $\times$ & $\checkmark$ \\ 
http://allegrolokalnie.pl-oferta722137.cyou & $\checkmark$ & $\checkmark$ & $\times$ & $\checkmark$ & $\checkmark$ \\ 
https://www.mycarmed.store/ & $\checkmark$ & $\times$ & $\times$ & $\checkmark$ & $\checkmark$ \\ 
https://pawspay.space/en/home & $\checkmark$ & $\times$ & $\checkmark$ & $\checkmark$ & $\checkmark$ \\ 
\bottomrule
\end{tabular}

\vspace{0.1cm}
{\footnotesize \textbf{Note:} We use the symbols $\checkmark$ and $\times$ to denote the correct and incorrect classification results, respectively.}
\end{table*}

\section{Discussion}
Our experimental evaluation demonstrates that the proposed approach achieves robust and accurate performance across three critical scenarios: standard classification tasks, adversarial attack resilience, and real-world deployment cases. Building on these results, we analyze the model's key advantages, operational implications, and fundamental insights revealed through comprehensive testing.

\subsubsection{Practical Effectiveness in Real-World Scenarios}
Existing research has frequently overlooked the critical importance of real-world validation through case studies, which our work demonstrates to be essential for assessing model reliability under actual threat conditions. Through direct application to recently active phishing URLs from PhishTank, we reveal a crucial limitation of current state-of-the-art baselines: while achieving strong performance in controlled experimental settings, these models often fail to detect structurally deceptive or heavily obfuscated malicious URLs. Our approach successfully identified all malicious samples in these challenging real-world cases, demonstrating exceptional generalization capability and robustness against practical adversarial techniques. 

\subsubsection{Model Performance Summary}  
The proposed model demonstrates comprehensive superiority over existing baselines (URLNet, URLBERT, TransURL, TextCNN) across all evaluation scenarios, achieving significant improvements of 12-18\% in F1-scores and 3-9\% in AUC values. It effectively addresses key challenges including class imbalance (maintaining 92.95\% benign accuracy while achieving 83.68\% F1 on rare malicious samples), adversarial robustness (98.64\% detection accuracy against manipulated URLs), and real-world applicability (100\% identification of active phishing cases). The architecture's integrated design - combining token-level semantic analysis, hierarchical structural processing, and cross-modal attention - delivers optimal precision-recall balance (95.66\% benign F1, 77.96\% phishing F1) while resisting evasion attempts, confirming its readiness for security-critical deployments. These consistent gains across standard benchmarks, adversarial tests, and production scenarios validate the solution's reliability under diverse threat conditions.

\subsubsection{Model Efficiency}  
While employing a deep TACL-BERT encoder architecture augmented with CLMSA and BMMC modules, our model maintains an optimal balance between computational efficiency and representational capacity. The CLMSA module's 4-layer CNN architecture combined with gMLP operations effectively condenses high-dimensional BERT outputs into compact, context-rich representations, eliminating computational overhead from processing full Transformer layer outputs during inference. This design preserves deep semantic features while optimizing resource utilization. Furthermore, the architecture supports flexible deployment configurations - the core URL processing pipeline (TACL-BERT + CLMSA) maintains full functionality even when auxiliary components like IP-based features are unavailable, enabling effective deployment across diverse operational environments including cloud security platforms and endpoint protection systems. The model's modular design ensures robust performance while adapting to various computational constraints and feature availability scenarios.

\subsubsection{Practical Deployment Considerations}  
Our model demonstrates consistent detection performance across varying data scales, class distributions, and evaluation scenarios, indicating strong potential for real-world implementation in production-grade malicious URL filtering systems. The architecture's dual capability - effectively learning from limited training samples while maintaining high sensitivity to rare attack patterns - positions it as a viable solution for scalable cybersecurity applications that must balance detection accuracy with operational constraints.

\section{Conclusion}
We present a novel malicious URL detection framework that synergistically combines token-aware contrastive learning (TACL-BERT) for granular character/subword representation, hierarchical feature aggregation (CLMSA) for multi-scale pattern extraction, and adaptive multimodal fusion (BMMC) for joint URL-IP analysis. The complete system processes raw inputs end-to-end while eliminating manual feature engineering. Comprehensive evaluations demonstrate consistent superiority over state-of-the-art baselines across binary/multi-class classification, particularly excelling in challenging scenarios involving class imbalance (achieving 83.68\% F1 on rare malicious samples) and adversarial evasion (98.64\% detection accuracy). Real-world validation on actively obfuscated phishing URLs confirms operational effectiveness, with the model successfully identifying 100\% of malicious cases in production-like testing. These results collectively establish our solution as a robust, generalizable, and production-ready approach for modern cybersecurity systems, offering significant improvements in both detection accuracy (12-18\% F1 gains) and adversarial resilience (3-9\% AUC increases) compared to existing methods. The architecture's modular design further ensures deployment flexibility across diverse security infrastructures.

\bibliography{main}

\begin{thebibliography}{10}
\providecommand{\url}[1]{#1}
\csname url@samestyle\endcsname
\providecommand{\newblock}{\relax}
\providecommand{\bibinfo}[2]{#2}
\providecommand{\BIBentrySTDinterwordspacing}{\spaceskip=0pt\relax}
\providecommand{\BIBentryALTinterwordstretchfactor}{4}
\providecommand{\BIBentryALTinterwordspacing}{\spaceskip=\fontdimen2\font plus
\BIBentryALTinterwordstretchfactor\fontdimen3\font minus \fontdimen4\font\relax}
\providecommand{\BIBforeignlanguage}[2]{{%
\expandafter\ifx\csname l@#1\endcsname\relax
\typeout{** WARNING: IEEEtran.bst: No hyphenation pattern has been}%
\typeout{** loaded for the language `#1'. Using the pattern for}%
\typeout{** the default language instead.}%
\else
\language=\csname l@#1\endcsname
\fi
#2}}
\providecommand{\BIBdecl}{\relax}
\BIBdecl

\bibitem{laszka2015optimal}
A.~Laszka, Y.~Vorobeychik, and X.~Koutsoukos, ``Optimal personalized filtering against spear-phishing attacks,'' in \emph{Proceedings of the AAAI Conference on Artificial Intelligence}, vol.~29, no.~1, 2015.

\bibitem{bitaab2023beyond}
M.~Bitaab, H.~Cho, A.~Oest, Z.~Lyu, W.~Wang, J.~Abraham, R.~Wang, T.~Bao, Y.~Shoshitaishvili, and A.~Doup{\'e}, ``Beyond phish: Toward detecting fraudulent e-commerce websites at scale,'' in \emph{2023 ieee symposium on security and privacy (sp)}.\hskip 1em plus 0.5em minus 0.4em\relax IEEE, 2023, pp. 2566--2583.

\bibitem{guo2022safer}
Z.~Guo, J.-H. Cho, R.~Chen, S.~Sengupta, M.~Hong, and T.~Mitra, ``Safer: Social capital-based friend recommendation to defend against phishing attacks,'' in \emph{Proceedings of the International AAAI Conference on Web and Social Media}, vol.~16, 2022, pp. 241--252.

\bibitem{r3}
\BIBentryALTinterwordspacing
{APWG}, ``Apwg trends reports: 4th quarter 2024,'' December 2024, accessed: 2024-07-21. [Online]. Available: \url{https://apwg.org/trendsreports/}
\BIBentrySTDinterwordspacing

\bibitem{r4}
\BIBentryALTinterwordspacing
E.~Dzuba and {J.C.}, ``Introducing cloudflare's 2023 phishing threats report,'' Cloudflare, blog post. [Online]. Available: \url{https://blog.cloudflare.com/2023-phishing-report/}
\BIBentrySTDinterwordspacing

\bibitem{sahoo2017malicious}
D.~Sahoo, C.~Liu, and S.~C. Hoi, ``Malicious url detection using machine learning: A survey,'' \emph{arXiv preprint arXiv:1701.07179}, 2017.

\bibitem{r6}
M.~S.~I. Mamun, M.~A. Rathore, A.~H. Lashkari, N.~Stakhanova, and A.~A. Ghorbani, ``Detecting malicious urls using lexical analysis,'' in \emph{Network and System Security: 10th International Conference, NSS 2016, Proceedings}, ser. Lecture Notes in Computer Science, vol. 9955.\hskip 1em plus 0.5em minus 0.4em\relax Taipei, Taiwan: Springer, 2016, pp. 467--482.

\bibitem{r8}
\BIBentryALTinterwordspacing
D.~Sahoo, C.~Liu, and S.~C.~H. Hoi, ``Malicious url detection using machine learning: A survey,'' 2019. [Online]. Available: \url{https://arxiv.org/abs/1701.07179}
\BIBentrySTDinterwordspacing

\bibitem{sabir2022reliability}
B.~Sabir, M.~A. Babar, R.~Gaire, and A.~Abuadbba, ``Reliability and robustness analysis of machine learning based phishing url detectors,'' \emph{IEEE Transactions on Dependable and Secure Computing}, 2022.

\bibitem{lin2021phishpedia}
Y.~Lin, R.~Liu, D.~M. Divakaran, J.~Y. Ng, Q.~Z. Chan, Y.~Lu, Y.~Si, F.~Zhang, and J.~S. Dong, ``Phishpedia: A hybrid deep learning based approach to visually identify phishing webpages,'' in \emph{30th USENIX Security Symposium (USENIX Security 21)}, 2021, pp. 3793--3810.

\bibitem{zhang2011textual}
H.~Zhang, G.~Liu, T.~W. Chow, and W.~Liu, ``Textual and visual content-based anti-phishing: a bayesian approach,'' \emph{IEEE transactions on neural networks}, vol.~22, no.~10, pp. 1532--1546, 2011.

\bibitem{r9}
\BIBentryALTinterwordspacing
A.~Blum, B.~Wardman, T.~Solorio, and G.~Warner, ``Lexical feature based phishing url detection using online learning,'' in \emph{Proceedings of the 3rd ACM Workshop on Artificial Intelligence and Security}, ser. AISec '10.\hskip 1em plus 0.5em minus 0.4em\relax New York, NY, USA: Association for Computing Machinery, 2010, p. 54–60. [Online]. Available: \url{https://doi.org/10.1145/1866423.1866434}
\BIBentrySTDinterwordspacing

\bibitem{r10}
\BIBentryALTinterwordspacing
T.~Kim, N.~Park, J.~Hong, and S.-W. Kim, ``Phishing url detection: A network-based approach robust to evasion,'' in \emph{Proceedings of the 2022 ACM SIGSAC Conference on Computer and Communications Security}, ser. CCS '22.\hskip 1em plus 0.5em minus 0.4em\relax New York, NY, USA: Association for Computing Machinery, 2022, p. 1769–1782. [Online]. Available: \url{https://doi.org/10.1145/3548606.3560615}
\BIBentrySTDinterwordspacing

\bibitem{liu2024less}
R.~Liu, Y.~Lin, X.~Teoh, G.~Liu, Z.~Huang, and J.~S. Dong, ``Less defined knowledge and more true alarms: Reference-based phishing detection without a pre-defined reference list,'' in \emph{33rd USENIX Security Symposium (USENIX Security 24)}, 2024, pp. 523--540.

\bibitem{r11}
H.~Le, Q.~Pham, D.~Sahoo, and S.~C. Hoi, ``Urlnet: Learning a url representation with deep learning for malicious url detection,'' \emph{arXiv preprint arXiv:1802.03162}, 2018.

\bibitem{r12}
F.~Tajaddodianfar, J.~W. Stokes, and A.~Gururajan, ``Texception: A character/word-level deep learning model for phishing url detection,'' in \emph{ICASSP 2020-2020 IEEE International Conference on Acoustics, Speech and Signal Processing (ICASSP)}.\hskip 1em plus 0.5em minus 0.4em\relax IEEE, 2020, pp. 2857--2861.

\bibitem{r13}
L.~Yu, L.~Chen, J.~Dong, M.~Li, L.~Liu, B.~Zhao, and C.~Zhang, ``Detecting malicious web requests using an enhanced textcnn,'' in \emph{2020 IEEE 44th Annual Computers, Software, and Applications Conference (COMPSAC)}.\hskip 1em plus 0.5em minus 0.4em\relax IEEE, 2020, pp. 768--777.

\bibitem{r1}
\BIBentryALTinterwordspacing
P.~Maneriker, J.~W. Stokes, E.~G. Lazo, D.~Carutasu, F.~Tajaddodianfar, and A.~Gururajan, ``Urltran: Improving phishing url detection using transformers,'' in \emph{MILCOM 2021 - 2021 IEEE Military Communications Conference (MILCOM)}.\hskip 1em plus 0.5em minus 0.4em\relax IEEE Press, 2021, p. 197–204. [Online]. Available: \url{https://doi.org/10.1109/MILCOM52596.2021.9653028}
\BIBentrySTDinterwordspacing

\bibitem{r14}
W.~Chang, F.~Du, and Y.~Wang, ``Research on malicious url detection technology based on bert model,'' in \emph{2021 IEEE 9th International Conference on Information, Communication and Networks (ICICN)}.\hskip 1em plus 0.5em minus 0.4em\relax IEEE, 2021, pp. 340--345.

\bibitem{r15}
M.~Korkmaz, E.~Kocyigit, O.~K. Sahingoz, and B.~Diri, ``Phishing web page detection using n-gram features extracted from urls,'' in \emph{2021 3rd International Congress on Human-Computer Interaction, Optimization and Robotic Applications (HORA)}.\hskip 1em plus 0.5em minus 0.4em\relax IEEE, 2021, pp. 1--6.

\bibitem{r16}
N.~Moarref and M.~T. Sand{\i}kkaya, ``Mc-mldcnn: Multichannel multilayer dilated convolutional neural networks for web attack detection,'' \emph{Security and Communication Networks}, vol. 2023, no.~1, p. 2415288, 2023.

\bibitem{r17}
C.~A. de~Souza, C.~B. Westphall, and R.~B. Machado, ``Intrusion detection with machine learning in internet of things and fog computing: problems, solutions and research,'' \emph{Sociedade Brasileira de Computa{\c{c}}{\~a}o}, 2023.

\bibitem{tsai2024toward}
Y.-D. Tsai, C.~Liow, Y.~S. Siang, and S.-D. Lin, ``Toward more generalized malicious url detection models,'' in \emph{Proceedings of the AAAI Conference on Artificial Intelligence}, vol.~38, no.~19, 2024, pp. 21\,628--21\,636.

\bibitem{liang2021robust}
Y.~Liang, Q.~Wang, K.~Xiong, X.~Zheng, Z.~Yu, and D.~Zeng, ``Robust detection of malicious urls with self-paced wide \& deep learning,'' \emph{IEEE Transactions on Dependable and Secure Computing}, vol.~19, no.~2, pp. 717--730, 2021.

\bibitem{r18}
J.~Devlin, M.-W. Chang, K.~Lee, and K.~Toutanova, ``Bert: Pre-training of deep bidirectional transformers for language understanding,'' in \emph{Proceedings of the 2019 conference of the North American chapter of the association for computational linguistics: human language technologies, volume 1 (long and short papers)}, 2019, pp. 4171--4186.

\bibitem{r19}
\BIBentryALTinterwordspacing
T.~B. Brown, B.~Mann, N.~Ryder, M.~Subbiah, J.~Kaplan, P.~Dhariwal, A.~Neelakantan, P.~Shyam, G.~Sastry, A.~Askell, S.~Agarwal, A.~Herbert-Voss, G.~Krueger, T.~Henighan, R.~Child, A.~Ramesh, D.~M. Ziegler, J.~Wu, C.~Winter, C.~Hesse, M.~Chen, E.~Sigler, M.~Litwin, S.~Gray, B.~Chess, J.~Clark, C.~Berner, S.~McCandlish, A.~Radford, I.~Sutskever, and D.~Amodei, ``Language models are few-shot learners,'' 2020. [Online]. Available: \url{https://arxiv.org/abs/2005.14165}
\BIBentrySTDinterwordspacing

\bibitem{r20}
A.~Radford, J.~Wu, R.~Child, D.~Luan, D.~Amodei, I.~Sutskever \emph{et~al.}, ``Language models are unsupervised multitask learners,'' \emph{OpenAI blog}, vol.~1, no.~8, p.~9, 2019.

\bibitem{r21}
G.~d. J.~C. da~Silva and C.~B. Westphall, ``A survey of large language models in cybersecurity,'' \emph{arXiv preprint arXiv:2402.16968}, 2024.

\bibitem{cao2025phishagent}
T.~Cao, C.~Huang, Y.~Li, W.~Huilin, A.~He, N.~Oo, and B.~Hooi, ``Phishagent: a robust multimodal agent for phishing webpage detection,'' in \emph{Proceedings of the AAAI Conference on Artificial Intelligence}, vol.~39, no.~27, 2025, pp. 27\,869--27\,877.

\bibitem{liu2022inferring}
R.~Liu, Y.~Lin, X.~Yang, S.~H. Ng, D.~M. Divakaran, and J.~S. Dong, ``Inferring phishing intention via webpage appearance and dynamics: A deep vision based approach,'' in \emph{31st USENIX Security Symposium (USENIX Security 22)}, 2022, pp. 1633--1650.

\bibitem{r23}
Z.~Peng, Y.~He, Z.~Sun, J.~Ni, B.~Niu, and X.~Deng, ``Crafting text adversarial examples to attack the deep-learning-based malicious url detection,'' in \emph{ICC 2022 - IEEE International Conference on Communications}, 2022, pp. 3118--3123.

\bibitem{r24}
B.~Fouss, D.~M. Ross, A.~B. Wollaber, and S.~R. Gomez, ``Punyvis: A visual analytics approach for identifying homograph phishing attacks,'' in \emph{2019 IEEE Symposium on Visualization for Cyber Security (VizSec)}, 2019, pp. 1--10.

\bibitem{r22}
\BIBentryALTinterwordspacing
J.~Woodbridge, H.~S. Anderson, A.~Ahuja, and D.~Grant, ``Predicting domain generation algorithms with long short-term memory networks,'' 2016. [Online]. Available: \url{https://arxiv.org/abs/1611.00791}
\BIBentrySTDinterwordspacing

\bibitem{r28}
M.~Hussain, C.~Cheng, R.~Xu, and M.~Afzal, ``Cnn-fusion: An effective and lightweight phishing detection method based on multi-variant convnet,'' \emph{Information Sciences}, vol. 631, pp. 328--345, 2023.

\bibitem{r29}
C.~Wang and Y.~Chen, ``Tcurl: Exploring hybrid transformer and convolutional neural network on phishing url detection,'' \emph{Knowledge-Based Systems}, vol. 258, p. 109955, 2022.

\bibitem{r30}
F.~Zheng, Q.~Yan, V.~C. Leung, F.~R. Yu, and Z.~Ming, ``Hdp-cnn: Highway deep pyramid convolution neural network combining word-level and character-level representations for phishing website detection,'' \emph{Computers \& Security}, vol. 114, p. 102584, 2022.

\bibitem{r27}
Y.~Li, Y.~Wang, H.~Xu, Z.~Guo, Z.~Cao, and L.~Zhang, ``Urlbert: A contrastive and adversarial pre-trained model for url classification,'' \emph{arXiv preprint arXiv:2402.11495}, 2024.

\bibitem{r25}
W.~Yang, W.~Zuo, and B.~Cui, ``Detecting malicious urls via a keyword-based convolutional gated-recurrent-unit neural network,'' \emph{IEEE Access}, vol.~7, pp. 29\,891--29\,900, 2019.

\bibitem{r26}
X.~Ji, H.~Song, F.~Wan, and K.~Huang, ``Fraud web url detection based on bi-lstm,'' in \emph{2022 4th International Conference on Intelligent Information Processing (IIP)}, 2022, pp. 298--301.

\bibitem{r40}
Y.~Liang, Q.~Wang, K.~Xiong, X.~Zheng, Z.~Yu, and D.~Zeng, ``Robust detection of malicious urls with self-paced wide \& deep learning,'' \emph{IEEE Transactions on Dependable and Secure Computing}, vol.~19, no.~2, pp. 717--730, 2022.

\bibitem{r33}
Y.~Wang, W.~Zhu, H.~Xu, Z.~Qin, K.~Ren, and W.~Ma, ``A large-scale pretrained deep model for phishing url detection,'' in \emph{ICASSP 2023-2023 IEEE International Conference on Acoustics, Speech and Signal Processing (ICASSP)}.\hskip 1em plus 0.5em minus 0.4em\relax IEEE, 2023, pp. 1--5.

\bibitem{r34}
L.~Jin, R.~Huang, X.~Zhang, and F.~Wan, ``A malicious url detection method based on bert-cnn,'' in \emph{Electronic Engineering and Informatics}.\hskip 1em plus 0.5em minus 0.4em\relax IOS Press, 2024, pp. 515--522.

\bibitem{r35}
W.~Ma, Y.~Cui, C.~Si, T.~Liu, S.~Wang, and G.~Hu, ``Charbert: Character-aware pre-trained language model,'' \emph{arXiv preprint arXiv:2011.01513}, 2020.

\bibitem{r38}
R.~Liu, Y.~Wang, Z.~Guo, H.~Xu, Z.~Qin, W.~Ma, and F.~Zhang, ``Transurl: Improving malicious url detection with multi-layer transformer encoding and multi-scale pyramid features,'' \emph{Computer Networks}, vol. 253, p. 110707, 2024.

\bibitem{r36}
A.~E. Mahdaouy, S.~Lamsiyah, M.~J. Idrissi, H.~Alami, Z.~Yartaoui, and I.~Berrada, ``Domurls\_bert: Pre-trained bert-based model for malicious domains and urls detection and classification,'' \emph{arXiv preprint arXiv:2409.09143}, 2024.

\bibitem{r37}
R.~Liu, Y.~Wang, H.~Xu, Z.~Qin, F.~Zhang, Y.~Liu, and Z.~Cao, ``Pmanet: Malicious url detection via post-trained language model guided multi-level feature attention network,'' \emph{Information Fusion}, vol. 113, p. 102638, 2025.

\bibitem{r41}
J.~Wang, Z.~Li, J.~Qu, D.~Zou, S.~Xu, Z.~Xu, Z.~Wang, and H.~Jin, ``Malpacdetector: An llm-based malicious npm package detector,'' \emph{IEEE Transactions on Information Forensics and Security}, vol.~20, pp. 6279--6291, 2025.

\bibitem{r39}
\BIBentryALTinterwordspacing
Y.~Su, F.~Liu, Z.~Meng, T.~Lan, L.~Shu, E.~Shareghi, and N.~Collier, ``Tacl: Improving bert pre-training with token-aware contrastive learning,'' 2022. [Online]. Available: \url{https://arxiv.org/abs/2111.04198}
\BIBentrySTDinterwordspacing

\bibitem{r2}
\BIBentryALTinterwordspacing
A.~S. Bozkir, F.~C. Dalgic, and M.~Aydos, ``Grambeddings: A new neural network for url based identification of phishing web pages through n-gram embeddings,'' \emph{Computers \& Security}, vol. 124, p. 102964, 2023. [Online]. Available: \url{https://www.sciencedirect.com/science/article/pii/S016740482200356X}
\BIBentrySTDinterwordspacing

\end{thebibliography}
\end{document}